\documentclass[aps,pre,twocolumn,superscriptaddress,floatfix]{revtex4}
\usepackage{graphicx}
\usepackage{bm}
\usepackage[dvips,unicode,colorlinks,linkcolor=blue,citecolor=blue,urlcolor=blue]{hyperref}

\begin{document}
\title{Semi-quantum molecular dynamics simulation of thermal properties and heat transport
in low-dimensional nanostructures}
\author{Alexander V. Savin}
\email[]{asavin@center.chph.ras.ru}
\affiliation{Semenov Institute of Chemical Physics, Russian Academy of Sciences,
Moscow 119991, Russia}

\author{Yuriy A. Kosevich}
\email[]{kosevich@polymer.chph.ras.ru,Yury.Kosevich@uv.es}
\affiliation{Semenov Institute of Chemical Physics, Russian Academy of Sciences,
Moscow 119991, Russia}
\affiliation{Materials Science Institute, University of Valencia, PO Box 22085, ES46071 Valencia, Spain}

\author{Andres Cantarero}
\email[]{Andres.Cantarero@uv.es}
\affiliation{Materials Science Institute, University of Valencia, PO Box 22085, ES46071 Valencia, Spain}

\date{\today}

\begin{abstract}
We present a detailed description of semi-quantum molecular dynamics
simulation of stochastic dynamics of a system of interacting particles. Within this approach, the dynamics
of the system is described with the use of classical Newtonian equations of motion in which the effects
of phonon quantum statistics are introduced through random Langevin-like forces with a specific
power spectral density (the color noise). The color noise describes the interaction of the molecular
system with the thermostat. We apply this technique to the simulation of thermal properties and heat
transport in different low-dimensional nanostructures. We describe the determination of
temperature in quantum lattice systems, to which the equipartition limit is not applied.
We show that one can determine the temperature of such system from the measured power spectrum
and temperature- and relaxation-rate-independent density of vibrational (phonon) states.
We simulate the specific heat and heat transport in carbon nanotubes, as well as the
heat transport in molecular nanoribbons with perfect (atomically smooth) and rough (porous) edges,
and in nanoribbons with strongly anharmonic periodic interatomic potentials. We show that
the effects of quantum statistics of phonons are essential for the carbon nanotube in the whole
temperature range $T<500$K, in which the values of the specific heat and thermal conductivity
of the nanotube are considerably less
than that obtained within the description based on classical statistics of phonons.
This conclusion is also applicable to other carbon-based materials and systems with high Debye
temperature like graphene, graphene nanoribbons, fullerene, diamond, diamond nanowires etc.
We show that the existence
of rough edges and quantum statistics of phonons change drastically the low temperature thermal
conductivity of the nanoribbon in comparison with that of the nanoribbon with perfect edges and classical
phonon dynamics and statistics. The semi-quantum molecular dynamics approach allows us to model the transition in the rough-edge nanoribbons from the thermal-insulator-like behavior at high temperature, when the thermal conductivity decreases with the conductor length, to the ballistic-conductor-like behavior at low temperature, when the thermal conductivity increases with the conductor length. We also show how the combination of strong nonlinearity
of periodic interatomic potentials with the quantum statistics of phonons changes completely
the low-temperature thermal conductivity of the system.
\end{abstract}

\maketitle
\section{Introduction \label{s0}}

Molecular dynamics (MD) is a method of numerical modeling of molecular systems based on classical Newtonian
mechanics. It does not allow for the description of pure quantum effects such as freezing out of high-frequency
oscillations at low temperatures and the related decrease to zero of heat capacity for
$T\rightarrow 0$. In classical molecular dynamics (CMD),
each dynamical degree of freedom possesses the same
energy $k_B T$, where $k_B$ is Boltzmann constant. Therefore, in classical statistics the specific heat
of a solid almost does not depend on temperature when only relatively small changes, caused the anharmonicity
of the potential, can be taken into account \cite{landau1}. On the other hand, because of its complexity,
a pure quantum-mechanical description does not allow in general the detailed modeling of the dynamics
of many-body systems. To overcome these obstacles, different semiclassical methods, which allow to take
into an account quantum effects in the dynamics of molecular systems, have been proposed
\cite{wang,donadio,heatwole,buyukdagli,ceriotti,dammak,wang2}.
The most convenient for the numerical modeling is the use of the Langevin equations
with color-noise random forces \cite{buyukdagli,dammak}.
In this approximation, the dynamics
of the system is described with the use of classical Newtonian equations of motion, while the
quantum effects are introduced through  random Langevin-like forces with a specific power spectral
density (the color noise), which describe the interaction of the molecular system with the thermostat.
Below we give a detailed description of this semi-quantum molecular dynamics (SQMD) approach in application
to the simulation of specific heat and heat transport in different low-dimensional nanostructures.

The paper is organized as follows. In Section \ref{s1} we describe the temperature-dependent Langevin
dynamics of the system under the action of random forces. If the  random forces are delta-correlated
in time domain, this corresponds to the white noise with a flat power spectral density. This situation
corresponds to high enough temperatures, when $k_B T$ is larger that the quantum of the highest-frequency
mode in the system, $k_B T\gg\hbar\omega_m$. But for low enough temperature, $k_B T\ll\hbar\omega_m$,
the stochastic dynamics of the system should be described with the use of random Langevin-like forces
with a non-flat power spectral density, which corresponds to the system with color noise.
In Section \ref{s2}  we describe the determination of temperature in quantum lattice systems,
to which the equipartition limit is not applied.
We show how one can determine the temperature of such system from the measured power spectrum
and temperature- and relaxation-rate-independent density of vibrational (phonon) states.
In Section \ref{s3} we describe a method for the generation of color noise with the power
spectrum, which is consistent with the quantum
fluctuation-dissipation theorem \cite{landau1,callen}.
In Sections \ref{s4} and \ref{s6} we apply such semi-quantum molecular dynamics approach
to the simulation of the specific
heat and phonon heat transport in carbon nanotubes. In Section \ref{s5} we apply this method
to the simulation of heat transport
in a molecular nanoribbon with rough edges. Previously, we have predicted analytically \cite{kosevich1} and confirmed by classical molecular dynamics simulations \cite{kosevich2} that
rough edges of molecular nanoribbon (or nanowire) cause strong suppression of phonon thermal
conductivity due to strong {\it momentum-nonconserving} phonon scattering. Here we show that the rough edges and quantum statistics
of phonons change drastically the thermal conductivity of the rough-edge nanoribbon in comparison
with that of the nanoribbon with perfect (atomically smooth) edges. The semi-quantum molecular dynamics approach allows us to model the transition in the rough-edge nanoribbons from the thermal-insulator-like behavior at high temperature, when the thermal conductivity decreases with the conductor length,  see Ref. \cite{kosevich2}, to the ballistic-conductor-like behavior at low temperature, when the thermal conductivity increases with the conductor length. In Section \ref{s7} we apply
this technique to the simulation of thermal transport in a nanoribbon with strongly anharmonic periodic
interatomic potentials. We show how the combination of strong nonlinearity of the interatomic
potentials with quantum statistics of phonons changes completely the
low-temperature thermal conductivity of the nanoribbon.
In Section \ref{s8} we provide a summary and discussions of all the main results of the paper.

\section{Langevin equations with color noise \label{s1}}

In the presence of a Langevin thermostat, equations  of motion of coupled particles have the
following form:
\begin{equation}
M_n\ddot{\rm\bf r}_n=-\frac{\partial}{\partial{\rm\bf r}_n}H-M_n\Gamma\dot{\rm\bf r}_n+{\Xi}_n,
\label{f1}
\end{equation}
where the three-dimensional radius-vector ${\rm\bf r}_n(t)$ gives the position of the $n$-th particle
at the time instant $t$ ($n=1,2,\dots,N$, $N$ being the total number of particles), $M_n$ is the particle
mass, $H$ is the Hamiltonian of the system and  ${\rm\bf F}_n=-\partial H/\partial{\rm\bf r}_n$
gives the force applied
to the $n$-th particle caused by the interaction with the other particles, $\Gamma=1/t_r$ is the friction
coefficient ($t_r$ being the relaxation time due to the interaction with the thermostat), and
$\Xi_n=\{\xi_{\alpha,n}\}_{\alpha=1}^3$ are random forces with the Gaussian distribution.

In the semi-quantum approach the random forces do not represent in general white noise. The power
spectral density of the random forces in that description should be given by the quantum
fluctuation-dissipation theorem \cite{landau1,callen}:
\begin{equation}
\langle\xi_{\alpha n}\xi_{\beta m}\rangle_\omega = 2M_n\Gamma k_BT\delta_{\alpha\beta}\delta_{nm} p(\omega,T),
\label{f2}
\end{equation}
where $n,m =1,2,...,N$, $\alpha,\beta=1,2,3$ are Cartesian components,
$\delta_{nm}$ and $\delta_{\alpha\beta}$ are Kronecker delta-symbols.
Here the positive and even in frequency $\omega$ function
\begin{eqnarray}
p(\omega, T)=
\frac12\frac{\hbar\omega}{k_BT}+\frac{\hbar\omega/k_BT}{\exp(\hbar\omega/k_BT)-1}
\label{f3}\\
=\frac{\hbar\omega}{2k_BT}\coth(\frac{\hbar\omega}{2k_BT})
\nonumber
\end{eqnarray}
gives the dimensionless power spectral density at temperature $T$ of an oscillator with
frequency $\omega$. The first term in Eq. (\ref{f3}) gives the contribution of the zero-point
oscillations to the power spectral density of random forces. With the use of Eqs. (\ref{f2}) and (\ref{f3}),
we get the correlation function of the random forces as
\begin{eqnarray}
\langle \xi_{\alpha n}(t)\xi_{\beta m}(0)\rangle
=\int_{-\infty}^{+\infty}\langle\xi_{\alpha n}\xi_{\beta m}\rangle_\omega
e^{i\omega t}\frac{d\omega}{2\pi}\nonumber \\
=2M_n\Gamma k_BT \delta_{\alpha\beta} \delta_{nm}
\int_{-\infty}^{+\infty}p(\omega,T)e^{i\omega t}\frac{d\omega}{2\pi}~.
\label{f4}
\end{eqnarray}

Let $\Omega_{max}$ be the maximal vibrational eigenfrequency of the molecular system. At high temperatures,
$T\gg\hbar\Omega_{max}/k_B$, one has $p(\omega,T)=1$ and the molecular system will perceive the random
forces as a delta-correlated white noise:
\begin{equation}
\langle \xi_{\alpha n}(t)\xi_{\beta m}(0)\rangle=2M_n\Gamma k_BT \delta_{\alpha\beta}\delta_{nm}\delta(t).
\label{f5}
\end{equation}
Therefore for high enough temperatures we deal with  molecular dynamics with Langevin uncorrelated
random forces. But for low temperatures, the random forces are correlated according to the function
given by Eqs. (\ref{f3}) and (\ref{f4}). In other words, for low temperatures the random forces in
the system represent the color noise with the dimensionless power spectral density $p(\omega,T)$
given by Eq. (\ref{f3}).

\section{Determination of temperature in quantum lattice systems\label{s2}}

In connection with the consideration of the thermodynamics and dynamics of non-classical lattices,
in this Section we discuss possible determination of the temperature in the systems, to which the
equipartition limit is not applied. We start from Eqs. (\ref{f1}), rewritten as the equations
of motion for the lattice vibrational mode with eigenfrequency $\Omega_n$:
\begin{equation}
M_n\ddot{\rm\bf Q}_n+M_n\Gamma\dot{\rm\bf Q}_n+M_n\Omega^2_n {\rm\bf Q}_n={\Xi}_n,
\label{f023}
\end{equation}
where $\Xi_n=\{\xi_{\alpha,n}\}_{\alpha=1}^3$ are random forces with the spectral
density given by Eqs.  (\ref{f2}) and (\ref{f3}).

From Eqs. (\ref{f2}), (\ref{f3}) and (\ref{f023}) we obtain
\begin{equation}
M_n\langle Q^2_n\rangle_{\omega,T}=\frac{\hbar\omega}{2}\coth(\frac{\hbar\omega}{2k_BT})\frac{2\Gamma}{(\Omega^2_n-\omega^2)^2+\omega^2\Gamma^2},
\label{f025}
\end{equation}

and power spectrum $\tilde{E}_T(\omega)$ of lattice vibrations with $3N-6$ nonzero eigenmodes:

\begin{eqnarray}
&&\tilde{E}_T(\omega)=\sum_{n=7}^{3N}\frac{1}{2}M_n(\omega^2 +\Omega_n^2) \langle Q^2_n\rangle_{\omega,T}\nonumber \\
&&=\sum_{n=7}^{3N}M_n\omega^2 \langle Q^2_n\rangle_{\omega,T}= \sum_{n=7}^{3N}M_n\Omega_n^2 \langle Q^2_n\rangle_{\omega,T}\nonumber \\
&&=\frac{\hbar\omega}{2}\coth(\frac{\hbar\omega}{2k_BT})\sum_{n=7}^{3N}\frac{\Gamma(\omega^2+\Omega_n^2)}{(\Omega^2_n-\omega^2)^2+\omega^2\Gamma^2}\nonumber\\
&&=\frac{\hbar\omega}{2}\coth(\frac{\hbar\omega}{2k_BT})\sum_{n=7}^{3N}\frac{2\Gamma\omega^2}{(\Omega^2_n-\omega^2)^2+\omega^2\Gamma^2}\nonumber \\
&&=\frac{\hbar\omega}{2}\coth(\frac{\hbar\omega}{2k_BT})\sum_{n=7}^{3N}\frac{2\Gamma\Omega_n^2}{(\Omega^2_n-\omega^2)^2+\omega^2\Gamma^2}.
\label{f026}
\end{eqnarray}

We can use $\langle Q^2_n\rangle_{\omega,T}$ and the power spectrum $\tilde{E}_T(\omega)$,  Eqs. (\ref{f025})
and (\ref{f026}), to compute the average energy of the system. Taking into account that the above
quantities are even functions of frequency $\omega$, for positive $\omega$ the average energy
of the system can be computed as:
\begin{equation}
E(T)=\int_{-\infty}^{\infty}\tilde{E}_T(\omega)\frac{d\omega}{2\pi}=\int_0^{\infty}\tilde{E}_T(\omega)\frac{d\omega}{\pi}.
\label{f030}
\end{equation}

In the weakly dissipative limit, with $\Gamma\ll\Omega_n$ for all $\Omega_n$, which is realized
in all the considered system, for positive $\omega$ and $\Omega_n$ one has, in the sense of generalized functions:
\begin{eqnarray}
&&\frac{2\Gamma\omega^2}{(\Omega^2_n-\omega^2)^2+\omega^2\Gamma^2}\approx\frac{2\Gamma\Omega_n^2}{(\Omega^2_n-\omega^2)^2+\omega^2\Gamma^2} \nonumber \\
\approx&&\frac{\Gamma(\Omega_n^2+\omega^2)}{(\Omega^2_n-\omega^2)^2+\omega^2\Gamma^2}\approx\pi\delta(\omega-\Omega_n).
\label{f031}
\end{eqnarray}

We can introduce in this limit the density of vibrational (phonon) states (for positive frequencies) as:
\begin{equation}
D(\omega)=\sum_{n=7}^{3N}\frac{2\Gamma\Omega_n^2}{(\Omega^2_n-\omega^2)^2+\omega^2\Gamma^2}\approx\pi\sum_{n=7}^{3N}\delta(\omega-\Omega_n).
\label{f032}
\end{equation}
Then the power spectrum (for positive $\omega$) consists of an array of the delta-functions
at the system eigenfrequencies, weighted by temperature,
\begin{eqnarray}
\tilde{E}_T(\omega)&\equiv&\frac{\hbar\omega}{2}\coth(\frac{\hbar\omega}{2k_BT})D(\omega)\nonumber\\
&=&\frac{\pi}{2}\hbar\omega\coth(\frac{\hbar\omega}{2k_BT})\sum_{n=7}^{3N}\delta(\omega-\Omega_n),
\label{f033}
\end{eqnarray}
and the average energy of the system (\ref{f030}) has the following form:
\begin{equation}
E(T)=\sum_{n=7}^{3N}\left[\frac{1}{2}\hbar\Omega_n
+ \frac{\hbar\Omega_n}{\exp(\hbar\Omega_n/k_BT)-1}\right ].
\label{f034}
\end{equation}
Here the first term in the sum gives the temperature-independent contribution of zero-point oscillations
to the energy of the system. As one can see from Eqs. (\ref{f032})-(\ref{f034}), in the weakly dissipative
limit the density of vibrational states, power spectrum and average energy of the system do not depend
explicitly on the relaxation rate $\Gamma$.

From Eq. (\ref{f033}) we get the "quantum definition" \ of temperature $T$ through the power
spectrum of lattice vibrations at the given frequency:
\begin{equation}
T=\hbar\omega/\left(k_B\ln\left[\frac{1+A_T(\omega)}{1-A_T(\omega)}\right]\right),
\label{f035}
\end{equation}
where
\begin{equation}
A_T(\omega)=\hbar\omega D(\omega)/(2\tilde{E}_T(\omega)).
\label{f036}
\end{equation}

In the case when one omits the zero-point contribution to the spectrum of the random forces,
namely considers
\begin{equation}
p(\omega, T)=\frac{\hbar\omega/k_BT}{\exp(\hbar\omega/k_BT)-1}
\label{f037}
\end{equation}
in Eqs. (\ref{f2}) and (\ref{f023}), the temperature can be determined from the following equation:
\begin{equation}
T=\hbar\omega/(k_B\ln\left[1+2A_T(\omega)\right]),
\label{f038}
\end{equation}
where in the definition of the function $A_T(\omega)$ in Eq. (\ref{f036}) the function $\tilde{E}_T(\omega)$, in contrast to function $\tilde{E}_T(\omega)$ in Eq. (\ref{f026}),  does not have now the zero-point contribution,
\begin{equation}
\tilde{E}_T(\omega)=\frac{\hbar\omega}{\exp(\hbar\omega/k_BT)-1}\sum_{n=7}^{3N}\frac{2\Gamma\Omega_n^2}{(\Omega^2_n-\omega^2)^2+\omega^2\Gamma^2},
\label{f039}
\end{equation}
and, correspondingly, one has
\begin{equation}
E(T)=\sum_{n=7}^{3N}\frac{\hbar\Omega_n}{\exp(\hbar\Omega_n/k_BT)-1}.
\label{f040}
\end{equation}

In the equipartition limit, which is realized for $T\gg\hbar\Omega_{max}/k_B$, one has
$\tilde{E}_T(\omega)=k_BTD(\omega)$, $A_T(\omega)\ll 1$,  and both Eqs. (\ref{f035}) and (\ref{f038})
turn into the identity $T=T$.

Therefore one can determine the temperature of the quantum lattice system from the measured power spectrum
$\tilde{E}_T(\omega)$ and temperature- and relaxation-rate-independent density of vibrational (phonon)
states $D(\omega)$.
In Fig.~\ref{fig06} we show the power spectrum, computed within the semi-quantum approach with
$\Gamma =1/t_r$,  $t_r=0.4$~ps, line 1, and in the limit, when the spectrum is given by an array
of the smoothed delta-functions, line 3. As one can see in this figure, the particular choice of
the (long enough) relaxation time $t_r=1/\Gamma$ indeed does not affect the presented results.
The determination of temperature, given by Eqs. (\ref{f035}) and (\ref{f038}),
is valid for weakly anharmonic systems,
in which the power spectrum of lattice vibrations is close to the one, given by Eq. (\ref{f026}).
Carbon-based materials like carbon nanotubes, graphene and graphene nanoribbons, and crystal structures
with stiff valence bonds belong to such systems. It is worth mentioning that the temperature,
given by  Eq. (\ref{f035}) or  Eq. (\ref{f038}), does not depend on frequency $\omega$ only in the case of equilibrium
harmonic lattice excitations, driven by random forces with power spectrum given by the
Bose-Einstein distribution, Eqs. (\ref{f2}) and (\ref{f3}).  The anharmonicity of the lattice can
change the power spectrum of the excitations, see Fig.~\ref{fig12} (b) below. In such case the
"phonon temperature" \ starts to depend on frequency (similar to the frequency- and
direction-dependent temperature of photons in non-equilibrium radiation, see Ref. \cite{landau1}),
and "hot phonons" \ with non-equilibrium high frequencies appear in the system,
see Section VI below.\

\section{Color noise generation \label{s3}}

For the implementation of the semi-quantum approach in molecular dynamics, random forces with the
power spectral density given by $p(\bar\omega)$ must be generated. Existing numerical techniques
allow the generation of a random quantity from a prescribed correlation function \cite{maradu}.
This approximation was used in \cite{dammak} in the modeling of thermodynamic
properties of liquid $^4$He above the $\lambda$ point. But universal methods require in general
the usage of high numerical resources. For instance, the utilization of the method proposed
in \cite{maradu} requires the execution of the Fast Fourier Transform for every random
force value. We propose a technique which takes into account the specific form of the
dimensionless power spectral density  $p(\bar\omega)$, where $\bar\omega=\hbar\omega/k_BT$ is a
dimensionless frequency.

In terms of  $\bar\omega$, the dimensionless power spectral density of random forces looks as
\begin{equation}
p(\bar\omega)=\frac12\bar\omega+\frac{\bar\omega}{\exp(\bar\omega)-1}=\frac12{\bar\omega}\coth(\bar\omega/2).
\label{f6}
\end{equation}

The random force with dimensionless power spectral density, given by Eq. (\ref{f6}), can be
presented as a sum of two independent functions, $p(\bar\omega)=p_0(\bar\omega)+p_1(\bar\omega)$,
where the first function, $p_0(\bar\omega)=\bar\omega/2$, has a linear power spectral density,
while the second one is $p_1(\bar\omega)=\bar\omega/[\exp(\bar\omega)-1]$.

To generate the color noise with a given dimensionless power spectral density $p(\bar\omega)$,
we will construct the dimensionless random vector functions
${\bf\rm S}_{n}(\tau)=\{S_{\alpha,n}\}_{\alpha=1}^3={\bf\rm S}_{0n}(\tau)+{\bf\rm S}_{1n}(\tau)$
of the dimensionless time $\tau=tk_BT/\hbar$, which will give the power spectral density
of the random forces in Eqs. (\ref{f1}) as
\begin{equation}
\langle\xi_{\alpha n}\xi_{\beta m}\rangle_\omega
=2M_n\Gamma k_BT\langle S_{\alpha n}S_{\beta m}\rangle_{\bar\omega},
\label{f7}
\end{equation}
such that
\begin{eqnarray}
&&\langle \xi_{\alpha n}(t)\xi_{\beta m}(0)\rangle
=\int_{-\infty}^{+\infty}\langle\xi_{\alpha n}\xi_{\beta m}\rangle_\omega
e^{i\omega t}\frac{d\omega}{2\pi}\nonumber \\
&&=2M_n\Gamma k_BT
\int_{-\infty}^{+\infty}\langle S_{\alpha n}S_{\beta m}\rangle_{\bar\omega}
e^{i\omega t}\frac{d\omega}{2\pi}\nonumber \\
&&=\frac{2}{\hbar}M_n\Gamma(k_BT)^2
\int_{-\infty}^{+\infty}\langle S_{\alpha n}S_{\beta m}\rangle_{\bar\omega}
e^{i\bar\omega\tau}\frac{d\bar\omega}{2\pi}~,
\label{f8}
\end{eqnarray}
and
\begin{equation}
\langle S_{\alpha n}S_{\beta m}\rangle_{\bar\omega}
=\langle S_{0\alpha n}S_{0\beta m}\rangle_{\bar\omega}+\langle S_{1\alpha n}S_{1\beta m}\rangle_{\bar\omega}.
\label{f9}
\end{equation}

Here, the uncorrelated random functions ${\bf\rm S}_{0n}(\tau)$ and ${\bf\rm S}_{1n}(\tau)$ will generate,
in a {\it finite frequency interval}, the power spectra $p_0(\bar\omega)$ and $p_1(\bar\omega)$, respectively.

For a molecular system with maximal vibrational eigenfrequency $\omega_m$, we will construct the
dimensionless random vector function ${\bf\rm S}_{0n}(\tau)$, whose power spectral density
generates $p_0(\bar\omega)$ in the frequency interval $[0,\bar\omega_m]$ ($\bar\omega_m=\hbar\omega_m/k_BT$).
Each component of this vector can be written as a sum of dimensionless random functions
\begin{equation}
S_{0\alpha n}(\tau)=\sum_{i=1}^4c_i[\eta_{\alpha n,i}(\tau)-\zeta_{\alpha n,i}(\tau)].
\label{f10}
\end{equation}
The functions $\{\zeta_{\alpha ni}\}_{i=1}^4$ which generate the color noise, are the solutions
of the linear equations:
\begin{equation}
\zeta'_{\alpha n i}(\tau)=\lambda_i[\eta_{\alpha n i}(\tau)-\zeta_{\alpha n i}(\tau)],
\label{f11}
\end{equation}
where $\zeta'_{\alpha n i}(\tau)$ is the derivative with respect to $\tau$. Here $\eta_{\alpha n i}(\tau)$
are dimensionless white-noise random forces, with the correlation functions
\begin{equation}
\langle\eta_{\alpha n i}(\tau)\eta_{\beta k j}(0)\rangle=2\delta_{\alpha\beta}\delta_{nk}\delta_{ij}\delta(\tau)/\lambda_i.
\label{f12}
\end{equation}

By solving Eqs. (\ref{f11}) in the frequency domain, $\zeta_{\alpha ni}$ can be substituted
in Eq. (\ref{f10}) and the power spectra of  ${\rm\bf S}_{0n}$ can be obtained as
\begin{equation}
\langle S_{0\alpha n}S_{0\beta m}\rangle_{\bar\omega}=
\delta_{\alpha\beta}\delta_{mn}\sum_{i=1}^4 \frac{2c_i^2\bar\omega^2}{\lambda_i(\lambda_i^2+\bar\omega^2)}.
\label{f13}
\end{equation}
The dimensionless parameters $c_i$ and $\lambda_i$ can be found by minimizing the integral
\begin{equation}
\int_0^1\left[\frac12x-\sum_{i=1}^4\frac{2\bar c_i^2x^2}{\bar\lambda_i(\bar\lambda_i^2+x^2)}\right]^2dx
\label{f14}
\end{equation}
with respect to the parameters $\bar c_1,\bar\lambda_1,...,\bar c_4,\bar\lambda_4$,
where  $\bar c_i=c_i/\bar\omega_m$, $\bar\lambda_i=\lambda_i/\bar\omega_m$.
The coefficients $c_i$, $\lambda_i$, $i=1,...,4$, obtained within this procedure,
are shown in Table \ref{tab1}. For these parameters, integral (\ref{f14})
reaches its lower value, equal to $2.46\times 10^{-8}$.

\begin{table}[tb]
\centering\noindent
\caption{Value of the coefficients $\lambda_i$, $c_i$, $\bar\Omega_i$, $\bar\Gamma_i$.
}
\label{tab1}
\begin{tabular}{cc|cc}
\hline
\hline
~~coefficient~~ & ~~~value~~~ & ~~~coefficient~~~ & ~~value~~\\
\hline
$\lambda_1/\bar\omega_m$ & 1.763817 & $c_5$ & 1.8315 \\
$\lambda_2/\bar\omega_m$ & 0.394613 & $c_6$ & 0.3429 \\
$\lambda_3/\bar\omega_m$ & 0.103506 & $\bar\Omega_5$ & 2.7189\\
$\lambda_4/\bar\omega_m$ & 0.015873 & $\bar\Omega_6$ & 1.2223\\
$c_1/\bar\omega_m$ & 1.043576 & $\bar\Gamma_5$ & 5.0142\\
$c_2/\bar\omega_m$ & 0.177222 & $\bar\Gamma_6$ & 3.2974\\
$c_3/\bar\omega_m$ & 0.050319 & ~ & ~\\
$c_4/\bar\omega_m$ & 0.010241 & ~ & ~\\
\hline
\hline
\end{tabular}
\end{table}

\begin{figure}[t]
\includegraphics[angle=0, width=1\linewidth]{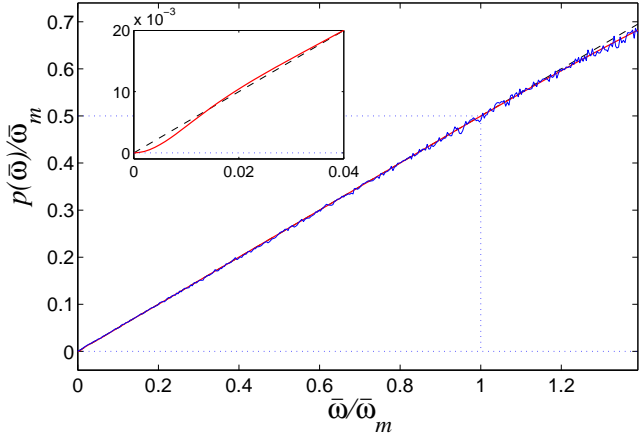}
\caption{(Color online)
Power spectral density of color noise $S_0(\tau)$ (blue line). Red line is given by Eq. (\ref{f13}),
dashed line shows the linear function $p_0(\bar\omega)/\bar\omega_m=\bar\omega/2\bar\omega_m$.
The two lines practically coincide in the frequency interval $0.015\le\bar\omega/\bar\omega_m\le1.1$.
The inset shows spectral densities given by Eq. (\ref{f13}) (red line) and by linear function
$p_0(\bar\omega)/\bar\omega_m$ (dashed line) in the frequency interval
$0\le\bar\omega/\bar\omega_m\le 0.04$. For color noise generation, Eqs. (\ref{f11}) were
numerically integrated with the use of fourth order Runge-Kutta method with a constant
integration step $\Delta \tau=0.05/\bar\omega_m$. }
\label{fig01}
\end{figure}

As one can see from Fig. \ref{fig01}, the function given by Eq. (\ref{f13}) approximates with
high accuracy the linear function
$p_0(\bar\omega)=\bar\omega/2$ in the frequency interval $0.015\le\bar\omega/\bar\omega_m<1.1$.

On the other hand, the random function $S_{1\alpha n}(\tau)$, which will generate the power
spectral density $p_1(\bar\omega)$,
can be approximated by a sum of two random functions with relatively narrow frequency spectra:
\begin{equation}
S_{1\alpha n}(\tau)=c_5\zeta_{\alpha n5}(\tau)+c_6\zeta_{\alpha n6}(\tau).
\label{f15}
\end{equation}
In this sum the dimensionless random functions $\zeta_{\alpha ni}(\tau)$,  $i=5,6$, satisfy
the equations of motion as
\begin{equation}
\zeta''_{\alpha ni}(\tau)=\eta_{\alpha ni}(\tau)-\bar\Omega_i^2\zeta_{\alpha ni}(\tau)
-\bar\Gamma_i\zeta'_{\alpha ni}(\tau),
\label{f16}
\end{equation}
where, as before, $\eta_{\alpha ni}(\tau)$ are delta-correlated white noise functions:
\begin{equation}
\langle\eta_{\alpha ni}(\tau)\eta_{\beta kj}(0)\rangle=2\bar\Gamma_i\delta_{\alpha \beta}\delta_{nk}\delta_{ij}\delta(\tau).
\label{f17}
\end{equation}

We can solve, as previously, Eq. (\ref{f16}) in the frequency domain and insert $\zeta_{\alpha ni}$
into Eq. (\ref{f15}) to obtain the power spectrum of $S_{1\alpha n}$:
\begin{equation}
\langle S_{1\alpha n}S_{1\beta m}\rangle_{\bar\omega}
=\delta_{\alpha\beta}\delta_{mn}\sum_{i=5}^6 \frac{2c_i^2\bar\Gamma_i}{(\bar\Omega_i^2-\bar\omega^2)^2+\bar\omega^2\bar\Gamma_i^2}
\label{f18}
\end{equation}

The dimensionless parameters $\{c_i,\bar\Omega_i,\bar\Gamma_i\}_{i=5}^6$
can be found by minimizing the integral
\begin{eqnarray}
\int_0^{\infty}\left[\frac{\bar\omega}{\exp(\bar\omega)-1}- \sum_{i=5}^6
\frac{2c_i^2\bar\Gamma_i}{(\bar\Omega_i^2-\bar\omega^2)^2+\bar\omega^2\bar\Gamma_i^2}\right]^2d\bar\omega
\label{f19}
\end{eqnarray}
with respect to the parameters $c_5,\bar\Gamma_5,\bar\Omega_5,c_6,\bar\Gamma_6,\bar\Omega_6$.
The obtained parameters are shown in Table \ref{tab1}. For these parameters, integral (\ref{f19})
reaches its lower value, equal to $2.3\times 10^{-4}$.
Since the integral (\ref{f19}) is determined by the
rapid-descending functions, the infinite integration limit $[0,\infty)$ can be replaced by a
finite one $[0,\bar\omega_m]$. Numerical integration of the integral (\ref{f19}) shows that
its minimal value practically does not depend on the upper integration limit for $\bar\omega_m>20$.
\begin{figure}[t]
\includegraphics[angle=0, width=1\linewidth]{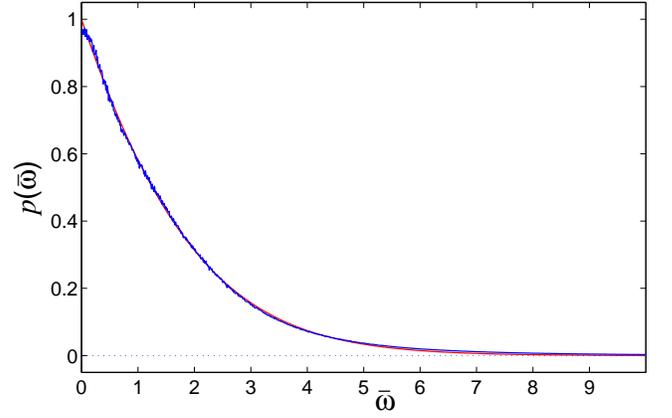}
\caption{(Color online)
Power spectral density of color noise $S_1(\tau)$, Eq. (\ref{f18}) (blue line).
Red line shows the function $p_1(\bar\omega)=\bar\omega/[\exp(\bar\omega)-1]$.
For color noise generation, Eqs. (\ref{f16}) were numerically integrated
with the use of fourth order Runge-Kutta method with a constant integration
step $\Delta\tau=0.02$. }
\label{fig02}
\end{figure}
\begin{figure}[t]
\includegraphics[angle=0, width=1\linewidth]{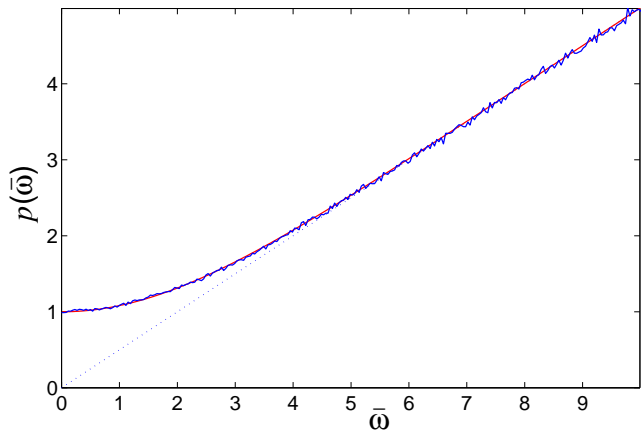}
\caption{(Color online)
Power spectral density of color noise
$S_0(\tau)+S_1(\tau)$, given by the sum of Eqs. (\ref{f13}) and (\ref{f18})  (blue line).
Red line shows the function $p(\bar\omega)=(\bar\omega/2)\coth(\bar\omega/2)$.
For color noise generation, Eqs. (\ref{f11}) and (\ref{f16}) were numerically integrated
with the use of fourth order Runge-Kutta method with a constant integration step
$\Delta t=0.005$. Maximal dimensionless frequency $\bar\omega_m=10$.
}
\label{fig03}
\end{figure}

As one can see from Fig. \ref{fig02}, the function, given by Eq. (\ref{f18}), approximates
with high accuracy the function
$p_1(\bar\omega)=\bar\omega/[\exp(\bar\omega)-1]$
in the frequency interval $0\leq\bar\omega\leq 10$.

Therefore in the semi-quantum molecular dynamics approach one has to solve
the Langevin equations (\ref{f1}) with
random forces $\Xi_n=\{\xi_{\alpha n}\}_{\alpha=1}^3$, whose power spectral density is determined as:
\begin{eqnarray}
&&\langle\xi_{\alpha n}\xi_{\beta m}\rangle_\omega
=2M_n\Gamma k_BT[\langle S_{0\alpha n}S_{0\beta m}\rangle_{\bar\omega}
+\langle S_{1\alpha n}S_{1\beta m}\rangle_{\bar\omega}]\nonumber \\
&&=2M_n\Gamma k_BT\delta_{\alpha\beta}\delta_{mn}\left[\sum_{i=1}^4
\frac{2c_i^2\bar\omega^2}{\lambda_i(\lambda_i^2+\bar\omega^2)}\right. \nonumber \\
&&\left. +\sum_{i=5}^6 \frac{2c_i^2\bar\Gamma_i}{(\bar\Omega_i^2-\bar\omega^2)^2+\bar\omega^2\bar\Gamma_i^2}\right].~~~
\label{f20}
\end{eqnarray}
From Eq. (\ref{f8}) we get the relation between the {\it dimension}
$\xi_{\alpha n}(t)$ and {\it dimensionless} $S_{\alpha n}(\tau)$ random forces:
$\xi_{\alpha n}(t)=\xi_{\alpha n}(\hbar\tau/k_BT)
=k_BT\sqrt{2M_n\Gamma/\hbar}[S_{0\alpha n}(\tau)+S_{1\alpha n}(\tau)]$.

In Fig. \ref{fig03} we show the comparison of the power spectral density for the dimensionless
frequency $\bar\omega$, given by the sum of Eqs. (\ref{f13}) and (\ref{f18}),  with the function
$(\bar\omega/2)\coth(\bar\omega/2)$, Eq. (\ref{f6}). We see a very good coincidence
in the frequency interval $[0,\bar\omega_m]$ for $\bar\omega_m=10$.

To describe the dynamics of molecular system with an account for quantum statistics of molecular
vibrations but without an account for zero-point oscillations, one has to keep only the last
two terms,  with $i=5,6$,  in Eq. (\ref{f20}) and to solve numerically Eqs. (\ref{f16})
in order to simulate random forces in Eq. (\ref{f1}).
Equations (\ref{f16}) play the role of the filter, which filters out the
dimensionless high-frequency ("non-quantum") component of the white noise at low temperature.

It is worth mentioning in this connection that the considered semi-quantum approach permits one to
obtain the correct value of the energy of zero-point oscillations. A semi-quantum account
for zero-point oscillations in the modeling of the properties of liquid $^4$He above the $\lambda$
point has allowed to describe correctly the liquid state of the helium \cite{dammak}, while the
classical molecular dynamics description  (without an account for zero-point oscillations)
predicts the solid state of the helium at the same low temperature.

\section{Temperature dependence of specific heat of carbon nanotube \label{s4}}

\begin{figure}[t]
\includegraphics[angle=0, width=1\linewidth]{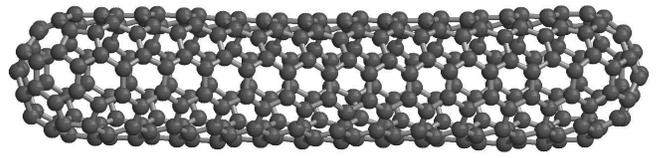}
\caption{ Form of a single-walled carbon nanotube C$_{300}$ with the zig-zag structure,
which consists of $N=300$ carbon atoms.}
\label{fig04}
\end{figure}

For the illustration of the implementation of the semi-quantum approach in molecular dynamics,
in this Section we find the temperature dependence of specific heat of a single-walled carbon
nanotube. We consider the nanotube with the (5,5) chirality index (the zig-zag structure),
which consists of $N=300$ carbon atoms, see Fig. \ref{fig04}. In the following we will use
the molecular-dynamics model of the carbon nanotube, which was discussed in details in Ref. \cite{savin1}.

To model nanotube dynamics at  temperature  $T$ in the classical approximation,
we will solve the Langevin equation (\ref{f1}) with a white noise, where $H$ is a Hamiltonian
of the nanotube and all $M_n\equiv M$, $M$ being a mass of a carbon atom.
The correlation functions of the random forces $\Xi_n$,
which describe the interaction of an $n$-th atom with the thermostat, are
given by Eq. (\ref{f5}).

\begin{figure}[t]
\includegraphics[angle=0, width=1\linewidth]{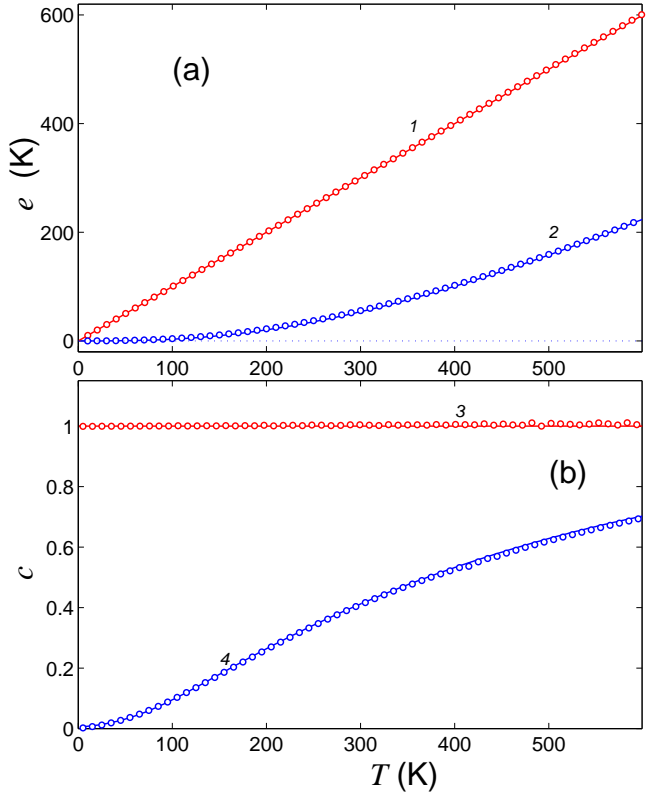}
\caption{(Color online)
Temperature dependence of the average thermal energy per mode $e=E/3Nk_B$ (a) and specific heat per mode $c=C/3Nk_B$ (b) of a single-walled carbon nanotube C$_{300}$. Lines 1 and 3 show  the dependencies, obtained with classical Langevin thermostat,
red circles give the result of numerical modeling.
Lines 2 and 4 give the dependencies, obtained within the quantum description with the use
of eigenmode spectra without an account for zero-point oscillations, Eqs.~(\ref{f040}) and (\ref{f22}),
blue circles give the result obtained within the semi-quantum molecular dynamics approach.}
\label{fig05}
\end{figure}

We take the initial conditions, which correspond to the equilibrium stationary state of the nanotube
at $T=0$, and integrate a system of equations of motion (\ref{f1}) for the time $t=20t_r$, during which
the system will reach the equilibrium state at finite temperature. The integration beyond this
time will allow us to find the average thermal energy of the system $\langle H\rangle=E(T)$. Then
the specific heat of the molecular system can be found from the temperature dependence of the average thermal
energy $C(T)=dE(T)/dT$. As one can see in Fig.~\ref{fig05}, the average thermal energy of the carbon nanotube, placed in classical Langevin thermostat, is strictly proportional to temperature, $E(T)=(3N-6)k_BT\approx 3Nk_BT$, line 1, and, correspondingly, nanotube specific heat almost
does not depend on temperature, line 3.
This result shows both the correctness of the modeling and that
carbon nanotubes are stiff structures which possess  weak anharmonicity of the
dynamics of the constituting atoms.
One can also see in Fig.~\ref{fig05} that the average thermal energy per mode $e(T)$ without an account for zero-point oscillations has the feature that $e(T)\leq k_B T$ in general and $e(T)\ll k_B T$ for temperature much less than the Debye one. For instance, in the carbon nanotube $e(T)$ is almost six times less than  $k_B T$ at room temperature $T$=300 K.

\begin{figure}[t]
\includegraphics[angle=0, width=1\linewidth]{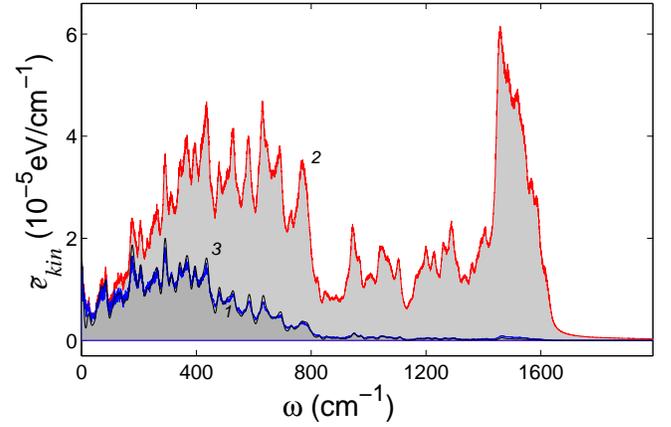}
\caption{(Color online)
Power spectral density of thermal vibrations of carbon nanotube C$_{300}$ atoms
at temperature $T=300$K. Lines 1 (blue) and 2 (red) give the densities obtained within
the semi-quantum and classical descriptions, respectively.
Line 3 (black) shows the density obtained with the use if Eq. (\ref{sa3}) for the system
of quantum oscillators, where the discreteness of the spectrum was smoothed for convenience.
The frequency spectrum of the mean kinetic energy per atom is shown.
}
\label{fig06}
\end{figure}

To obtain power spectra of the thermal oscillations of the atoms in the carbon nanotube, one has to
switch off the interaction with the thermostat after the establishment of the thermal equilibrium
in the system. In other words, one has to solve numerically the frictionless equations of motion,
Eqs.~(\ref{f1}) with $\Gamma=0$ and $\Xi_n={\rm\bf 0}$,
with the initial conditions for ${\rm\bf r}_n$ and $\dot{\rm\bf r}_n$, which correspond to the
thermalized state of the molecular system. By performing the Fast Fourier Transform of the
time-dependent particle velocities $\dot{\rm\bf r}_n(t)$, one can get the power spectrum
$\tilde{E}_T(\omega)$, Eq. (\ref{f026}). To increase the accuracy of the measurement,
it is necessary to
perform the averaging of the obtained results on different initial thermalized states
of the system. Power spectral density of the thermal vibrations of the carbon nanotube atoms
is shown in Fig.~\ref{fig06}. As one can see from this
figure, all vibrational modes of the system are excited in the classical Langevin thermostat.

Carbon nanotube is a rigid structure in which the anharmonicity of atomic dynamics at room
temperature is weakly pronounced: as one can see from line 2 in Fig.~\ref{fig06}, the
characteristic Debye temperature of the carbon nanotube is very high,
$T_{D}\sim\hbar\omega_m/k_B\approx 2400$K, where $\omega_m\approx 1640$~cm$^{-1}$ is the maximal phonon
frequency in the system. Therefore we can obtain the temperature dependence of the carbon nanotube
specific heat with the use of the spectrum of the harmonic (non-interacting)
vibrational eigenmodes: $C(T)=dE(T)/dT$,
where the average energy of the system $E(T)$ is given by Eq. (\ref{f034}).
To obtain the eigenfrequencies $\Omega_n$, we have to find all the eigenvalues of the square symmetric
matrix $\partial^2 H/\partial{\rm\bf r}_n\partial{\rm\bf r}_m$
of the $3N\times 3N$ size:  if $\lambda_n$ is one of the eigenvalues, the eigenfrequency is
determined as $\Omega_n=\sqrt{\lambda_n/M_n}$. For $N=300$ atoms in the nanotube, we have $3N=900$
eigenmodes, from which the first 6 modes, which describe rigid translations and rotations of the
nanotube, have zero frequency, $\Omega_1=\dots=\Omega_6=0$, while the other eigenfrequencies
are nonzero and are distributed in the interval
25.1~cm$^{-1}\le\Omega_n\le 1640.0$~cm$^{-1}$, $n=7,8,\dots,900$.
We can obtain the temperature dependence of the nanotube specific heat with the use of Eq. (\ref{f034}):
\begin{equation}
C(T)=k_B\sum_{n=7}^{3N}\left(\frac{\hbar\Omega_n}{k_BT}\right)^2 \frac{\exp(\hbar\Omega_n/k_BT)}{[\exp(\hbar\Omega_n/k_BT)-1]^2}
\label{f22}
\end{equation}

As one can see in Fig.~\ref{fig05}, the nanotube specific heat in the quantum description
monotonously goes to zero for
$T\rightarrow 0$, and the effects of quantum statistics of phonons are essential for the carbon
nanotube in the whole
temperature range $T<500$K, in which the specific heat $C(T)$ of the nanotube is considerably less
than the classical-statistics value $(3N-6)k_B$. This conclusion also applies to other carbon-based
materials and systems with high Debye temperature like graphene, graphene nanoribbons, fullerene,
diamond, diamond nanowires etc.

To model the nanotube stochastic dynamics in the semi-quantum approach, we will use the Langevin equations
of motion (\ref{f1}) with random forces $\Xi_n=\{\xi_{\alpha n}\}_{\alpha=1}^3$ with the power
spectral density, given by
$\langle\xi_{\alpha n}\xi_{\beta m}\rangle_\omega =2M\Gamma k_BT\langle S_{1\alpha n}S_{1\beta m}\rangle_{\bar\omega}$,
see Eq. (\ref{f20}).
Here it is explicitly taken into account that zero-point oscillations do not contribute to the
specific heat of the system, and therefore one needs to account only for the random functions,
given by Eq. (\ref{f18}).

Therefore the semi-quantum approach in this case amounts to the integration of Eqs. (\ref{f1})
and (\ref{f16}), to find the average energy  $\langle H\rangle=E(T)$ and the
specific heat $C(T)=dE(T)/dT$ of the system.
As one can see from Fig.~\ref{fig05}, the temperature
dependence of the carbon nanotube specific heat, obtained by studying the nanotube stochastic
dynamics with color noise, coincides completely with the dependence, obtained from the spectrum
of the harmonic vibrational eigenmodes in the system.
Power spectral density of carbon atoms vibrations, see Fig.~\ref{fig06}, shows
that the semi-quantum approach describes the thermalization of only the low-frequency eigenmodes, with
$\omega<\omega_T=k_BT/\hbar$. In such a way, the approach models the quantum freezing
of the high-frequency eigenmodes at low temperature $T<T_{D}$.  For the carbon nanotube, the
classical description is definitely not valid at room temperature $T=300$K, when the
characteristic frequency $\omega_T=208.5$~cm$^{-1}$ is much less than the maximal one in the system,
$\omega_m= 1640$~cm$^{-1}$, see Fig. \ref{fig06}.

It is worth mentioning in this connection that the obtained results for the temperature dependence
of the specific heat of the carbon nanotube C$_{300}$
almost do not depend on the relaxation time $t_r$. All the five values of $t_r$,
$t_r=0.1$, 0.2, 0.4, 0.8, 1.6~ps, give the same average energy $E(T)$ and specific heat $C(T)$
of the nanotube. One cannot use either too short $t_r$, when the system dynamics becomes
the forced one, or too long $t_r$, which increases the computation time. We find that the optimal
relaxation time for the carbon nanotube is $t_r=0.4$~ps.

Above we have introduced and discussed the lattice temperature of the system embedded in the Langevin
thermostat with color noise determined by the quantum fluctuation-dissipation theorem, see
Eqs. (\ref{f035}) and (\ref{f038}). But to compute the thermal conductivity, one also needs
to determine the local temperature of the part of the molecular system, which is placed
between two thermostats with different temperatures. In the classical-statistics approximation,
the lattice temperature can be determined as the average double kinetic energy per mode (or per degree of freedom):
\begin{equation}
T= \frac{1}{3Nk_B}\langle \sum_{n=1}^NM\dot{\bf r}_n^2\rangle,
\label{sa1}
\end{equation}
where $N$ is a number of atoms and is assumed that $3N-6\approx 3N$. As we have shown in
Chapter \ref{s2}, such determination of temperature does not work in the semi-quantum
approximation when one needs to deal with the density and population of vibrational (phonon) states.
In this case the determination of temperature is given by Eq. (\ref{f035}) or by Eq. (\ref{f038}).
But in the MD simulations, one can also use the determination of temperature, which is equivalent
to the integral presentation of Eq. (\ref{f033}).

Having in mind further simulation of thermal conductivity, we omit in Eqs. (\ref{f2}) and (\ref{f023})
the zero-point contribution to the spectrum of the random forces. With the use of Eqs. (\ref{f030}), (\ref{f033}),
(\ref{f039}) and (\ref{f040}), we find the average energy per mode (per degree of freedom) as
\begin{eqnarray}
&&e(T)\equiv\frac{1}{3N}E(T)=\frac{1}{3N}\int_0^\infty
\frac{\hbar\omega}{\exp(\hbar\omega/k_BT)-1}D(\omega)\frac{d\omega}{\pi}\nonumber\\
&&=\frac{1}{3N}\sum_{n=7}^{3N}\frac{\hbar\Omega_n}{\exp(\hbar\Omega_n/k_BT)-1},
\label{sa3}
\end{eqnarray}
where $D(\omega)$ is density of vibrational states, cf. Eq. (\ref{f032}).

From the MD simulations, we can also measure the spectral density $\tilde{e}_{kin}(\omega)$ of the average double kinetic energy
per mode $e_{kin}$ as
\begin{equation}
e_{kin}\equiv\frac{1}{3N}\sum_{n=1}^N\langle M_n\dot{\bf r}^2_n\rangle=
\int_0^\infty \tilde{e}_{kin}(\omega)\frac{d\omega}{\pi}.
\label{sa4}
\end{equation}

In the case of correct "quantum" \ level occupation, the both Eqs. (\ref{sa3}) and
(\ref{sa4}) should correspond to the same value of temperature,
therefore we obtain the following equation for the (numerical)  determination of the lattice temperature:
\begin{equation}
e(T)=e_{kin}.
\label{sa6}
\end{equation}
Since function $e(T)$ monotonously increases with $T$, Eq. (\ref{sa6}) has a unique solution
for the temperature. Similar equation, but with an account for zero-point oscillations, for the rescaling of MD temperature for an effective one was used in Ref. \cite{wang90} for quantum corrections of MD results. As we have mentioned above in discussion of Fig.~\ref{fig05}, the neglect of zero-point oscillations in the computation of the average thermal energy leads to the feature that $e_{kin}\leq k_B T$ in general and $e_{kin}\ll k_B T$ for temperature much less than the Debye one, see also Fig. {\ref{fig11} (b) and (c) below.

Essentially one can use Eq. (\ref{sa6}) for the determination of the lattice temperature only for
the quantum-statistics population of $all$ phonon states. But the situation can emerge when
high-frequency phonons have an excess excitation, in comparison with the Bose-Einstein
distribution for the given temperature, see section \ref{s6} below.
In this case the lattice temperature  $T$ can be determined with the use of only part
of the phonon spectrum:
\begin{equation}
\frac{1}{3N}\int_{\omega_1}^{\omega_2}\frac{\hbar\omega}{\exp(\hbar\omega/k_BT)-1}D(\omega)\frac{d\omega}{\pi}=
\int_{\omega_1}^{\omega_2}\tilde{e}_{kin}(\omega)\frac{d\omega}{\pi},
\label{sa5}
\end{equation}
where $[\omega_1,\omega_2]$ is the frequency interval with the Bose-Einstein distribution
of the average energy
per mode $\tilde{e}(\omega)$. For
$\omega_1=0$, $\omega_2>\Omega_{max}+\Gamma$ and  $k_BT\gg \hbar\omega_2$, Eq. (\ref{sa5}) gives the
classical definition of temperature, $e_{kin}=k_B T$. Clearly Eq. (\ref{sa6}), which is more
convenient for the numerical modeling,  gives the correct value of lattice temperature only
for $\omega_1=0$ and $\omega_2\ge \Omega_{max}+\Gamma$.

If the molecular system is completely embedded in the Langevin thermostat with color-noise random forces,
their spectrum, given by Eq. (\ref{f3}),  provides the correct "quantum" \ population of all the phonon states.
As one can see in Fig.~\ref{fig06}, at $T=300$K the spectral density of the average energy
per mode $\tilde{e}_{kin}(\omega)$ in the $C_{300}$  carbon nanotube (red line 1) coincides exactly with
the smoothed function $D(\omega)\hbar\omega/(\exp(\hbar\omega/k_BT)-1)/3N$ (black line 3).
In this case Eq. (\ref{sa6}) determines the temperature, which is exactly equal to the one
of the thermostat.

\section{Thermal conductivity of nanoribbon with rough edges \label{s5}}

First we apply the semi-quantum approach for the molecular dynamics simulation of thermal
conductivity in the nanoribbon with rough edges and harmonic interparticle potential. In such
system the vibrational eigenmodes do not interact and the considered semi-quantum approach
turns to be the exact one.

It was predicted analytically and confirmed by classical molecular dynamics simulations that
rough edges of molecular nanoribbon (or nanowire) cause strong suppression of phonon thermal
conductivity due to strong {\it momentum-nonconserving} phonon scattering \cite{kosevich1,kosevich2}.
In the case of nonlinear interatomic potential, the ribbon has a finite, length-independent
thermal conductivity, while in the case of harmonic interatomic potential,
the thermal conductivity decreases with the ribbon length (and the ribbon behaves
as a thermal insulator). Essentially for both interatomic potentials, the thermal conductivity
increases with the length of the nanoribbon with perfect (atomically smooth) edges.
In the case of harmonic interatomic potential, phonons in the nanoribbon with rough edges
experience effective localization due to strong antiresonance multichannel reflection from
side atomic oscillators in the rough edge layers (the Anderson-Fano-like localization due
to interference effects in phonon backscattering), see \cite{kosevich1,kosevich2}. Apparently
such strong backscattering can localize phonons with the wavelength shorter than the ribbon
length. Within the classical description, such effect does not depend on temperature. But
this picture will be changed if we take into account the quantum effects. The latter result
in thermalization and correspondingly in participation in low temperature thermal transport
of long wave phonons only. The long wave phonons are much less affected by surface roughness
than the short wave phonons and therefore the effect of thermal conductivity suppression
should decrease with the decrease of temperature. Below we demonstrate this effect
within the proposed semi-quantum approach.

We consider the system which consists of $K$ parallel molecular chains
in one plane \cite{kosevich2}. Let $k$ be the chain number, $n$ be the molecular number in the chain,
then the equilibrium position of the $n$-th atom in the $k$-th chain will be $x_{k,n}^0=na+[1+(-1)^k]a/4$,
$y_{k,n}^0=bk$, where $a$ and $b$ are, respectively, the intra- and interchain spatial periods,
see Fig.~\ref{fig07}.
\begin{figure}[tbp]
\includegraphics[angle=0, width=1\linewidth]{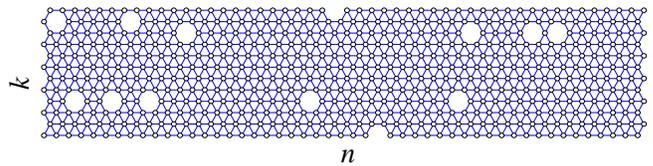}
\caption{
Molecular ribbon made of $K=12$ molecular chains.
Density and widths of rough edges are $d=0.95$ and $K_1=4$, respectively.
Lines connect the interacting atoms.
}
\label{fig07}
\end{figure}

Only the longitudinal displacements are taken into account in a simplest scalar model of
two-dimensional crystal:  $x_{k,n}(t)=x_{k,n}^0+au_{k,n}(t)$, $y_{k,n}\equiv y_{k,n}^0$,
where $u_{k,n}$ is a dimensionless longitudinal displacement of the $(k,n)$-th molecule from
its equilibrium position.
Hamiltonian of the system has the form as
\begin{eqnarray}
H&=&\sum_{k=1}^K\sum_{n=1}^N\frac12Ma^2\dot{u}_{kn}^2\nonumber\\
&+&E_0\sum_{k=1}^K\sum_{n=1}^{N-1}V(u_{k,n+1}-u_{k,n})
+E_0\sum_{k=1}^KE_k,
\label{f25}
\end{eqnarray}
where $N$ is a number of molecules in each chain,  $M$ is a mass of molecule, $E_0$ is a
characteristic interaction energy, $V(\rho)$ is the dimensionless potential of the nearest-neighbor
intrachain interactions, $\rho$ describes relative displacements of the nearest-neighbor atoms,
$E_k$ describes the interchain interaction.

Dimensionless energy  of the nearest-neighbor interchain interactions for odd $k$ is
\begin{equation}
E_k=\sum_{n=2}^N U(u_{k,n}-u_{k+1,n-1})+\sum_{n=1}^NU(u_{k+1,n}-u_{k,n}),
\label{f26}
\end{equation}
and for even $k$ is
\begin{equation}
E_k=\sum_{n=1}^N U(u_{k,n}-u_{k+1,n})+\sum_{n=1}^{N-1}U(u_{k+1,n+1}-u_{k,n}),
\label{f27}
\end{equation}
where $U(\rho)$ is the dimensionless potential of the nearest-neighbor interchain interaction.

We will consider a finite rectangular $(1\le n\le N,~1\le k\le K)$ with free edges,
and harmonic intra- and interchain interaction potentials:
\begin{equation}
V(r)=r^2/2,~~U(r)=r^2/4.
\label{f28}
\end{equation}

For the convenience of the modeling, we introduce the dimensionless quantities:
temperature $\tilde{T}=T/T_0$, time $\tilde{t}=t/t_0$, and energy $E=H/E_0$,
where $T_0=\hbar^2/Ma^2k_B$, $t_0=\hbar/k_BT_0$, and $E_0=Ma^2/t_0^2$.
Then the dimensionless Hamiltonian will have the form as
\begin{equation}
{\cal H}=\sum_{k=1}^K\sum_{n=1}^N\frac12\dot{u}_{m,n}^2+\sum_{k=1}^K\sum_{n=1}^{N-1}
V(u_{k,n+1}-u_{k,n})+\sum_{k=1}^{K}E_k,
\label{f29}
\end{equation}
where the dot denotes the derivative with respect to the dimensionless time $\tilde{t}$.
\begin{figure}[tbp]
\includegraphics[angle=0, width=1\linewidth]{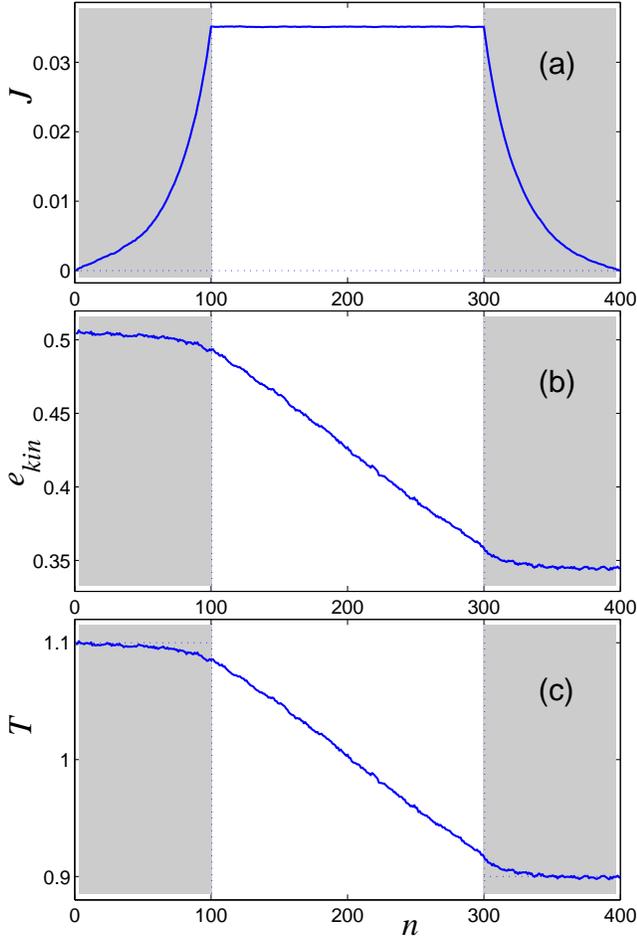}
\caption{(Color online)
Distribution of (a) local energy flux $J$, (b) local kinetic energy per mode $e_{kin}$
and (c) local temperature $T$ along the rough-edge ribbon
(ribbon length $N=400$, width $K=12$, rough edges widths $K_1=4$,
porosity of rough edges $p=0.05$).
Gray arias indicate the ribbon ends, embedded in semi-quantum Langevin
thermostats with temperatures $T_\pm=1\pm 0.1$.
}
\label{fig08}
\end{figure}

For the dimensionless quantities, equations of motion will take the form:
\begin{equation}
\ddot{u}_{kn}=-\frac{\partial\cal H}{\partial u_{kn}}-\tilde\Gamma\dot{u}_{kn}+\eta_{kn},~~k=1,...,K,~n=1,...,N
\label{f30}
\end{equation}
where $\tilde\Gamma=\Gamma t_0$ is the dimensionless friction, and the dimensionless random forces
$\eta_{kn}$ is the color noise with the correlation functions
\begin{equation}
\langle\eta_{kl}(s)\eta_{mn}(0)\rangle=2\tilde\Gamma\tilde{T}\delta_{km}\delta_{ln}
\int_{-\infty}^{+\infty}p(\tilde{\omega},\tilde{T})\exp(-i\tilde\omega s)\frac{d\tilde\omega}{2\pi},
\label{f31}
\end{equation}
and dimensionless power spectral density
\begin{equation}
p(\tilde\omega,\tilde{T})=
\frac12(\tilde\omega/\tilde{T})+\frac{\tilde\omega/\tilde{T}}{\exp(\tilde\omega/\tilde{T})-1},
\label{f32}
\end{equation}
where $\tilde\omega=\omega t_0$ is the dimensionless frequency.
In the following we will deal only with the dimensionless quantities.

Phonons do not interact in the system with harmonic interparticle potential. Therefore the
zero-point oscillations will not affect the thermal transport along the nanoribbon and can be
not taken into the account. In the latter case we can consider the dimensionless random forces
with the power spectra, given by
$\langle\eta_{kn}\eta_{lm}\rangle_{\tilde\omega}
=2\tilde{\Gamma}\tilde{T}\langle S_{1kn}S_{1lm}\rangle_{\bar\omega}$,
cf. Eq. (\ref{f20}). Random forces in Eq. (\ref{f30}) are $\eta_{kn}(\tilde{t})=\eta_{kn}(\tau/\tilde{T})
=\tilde{T}\sqrt{2\tilde{\Gamma}}S_{1 kn}(\tau)$.

Within this approach, we will describe the temperature dependence of the nanoribbon thermal
conductivity $\kappa(T)$.
We embed the left end of the ribbon with length $N_e=100$ in the thermostat with temperature $T_+=1.1T$,
while the right end embed in the thermostat with temperature $T_-=0.9T$. In this case the dynamics
of the atoms with numbers $n=1,...,N_e$ is described by Langevin equations (\ref{f30})
with temperature $T=T_+$, the atoms with numbers $n=N-N_e+1,...,N$ are described by these
equations with temperature $T=T_-$, while the atoms in the central part of the nanoribbon,
with numbers $n=N_e+1,...,N-N_e$, are described by the frictionless equations,
Eqs.  (\ref{f30}) with $\tilde\Gamma=0$ and $\eta_{kn}\equiv 0$.
We take the relaxation time $t_r=100$, at which the edge effects at the interfaces between
the ribbon and the thermostats are not essential. We integrate the equations of motion
to find the stationary energy flux $J$ along the ribbon.

To determine the temperature profile, first we find energy per mode
at the longitudinal coordinate $x=na$ by averaging the distribution across the ribbon
of the scalar-model double kinetic energy:
\begin{equation}
e_{kin}(n)=\frac{1}{K}\sum_{j=1}^K\langle \dot{u}^2_{n,j}\rangle.
\label{f32a}
\end{equation}
Since in the ribbon with a constant average width the phonon spectrum does not depend on the
longitudinal coordinate and the inter-mode interaction is absent in the harmonic system,
we can determine the local temperature from the equation $e_{kin}(n)=e(T_n)$, similar
to Eq. (\ref{sa6}).
Figure~\ref{fig08} shows the distribution along the ribbon of energy flux $J$, kinetic energy
per  mode $e_{kin}$ and temperature $T$. There is a stationary flux in the area between
the thermostats. Since the temperature distribution of phonon frequencies in the ribbon
with harmonic interactions is determined by the Langevin thermostats with Bose-Einstein
color noise, we can uniquely determine the local temperature $T_n$, which monotonously
changes from the value $T_+$ at the left end to the value  $T_-$ at the right end
with a linear gradient, see Fig.~\ref{fig08}(c).

Then the dimensionless thermal conductivity of the finite-length ribbon can be found as
\begin{equation}
\kappa(N,T)=(N-2N_e)J/[K(T_+-T_-)].
\label{f33}
\end{equation}

In the ribbon with perfect (atomically smooth) edges, phonons have infinite mean free path
and therefore the energy flux $J$ does not depend on the length of the central part of the
ribbon $N_c=N-2N_e$. In this case, as one can conclude from Eq. (\ref{f33}), the thermal
conductivity increases linearly with $N_c$ and therefore such ribbon behaves as an ideal
(ballistic) thermal conductor. In the absence of anharmonicity of interatomic interactions,
this conclusion is valid for any temperature.
\begin{figure}[tbp]
\includegraphics[angle=0, width=1\linewidth]{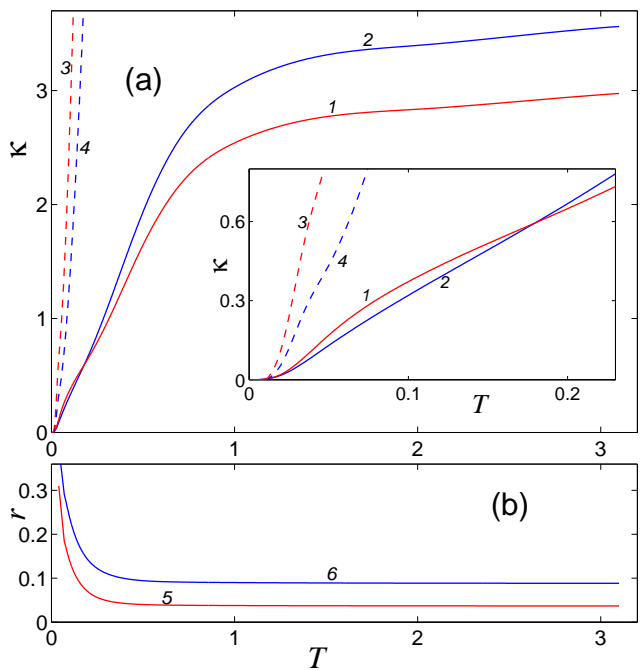}
\caption{(Color online)
(a) Thermal conductivity $\kappa$ of rough-edge ribbon (ribbon width $K=12$, rough edges widths $K_1=4$,
porosity of rough edges $p=0.05$) versus temperature $T$ for ribbon length $N=600$
(line 1) and  $N=400$ (line 2). Dashed lines 3 and 4 give the dependencies for perfect nanoribbons
(with zero porosity of edges, $p=0$) with $N=600$ and $N=400$, respectively. Inset shows the
low-temperature limit, $T\leq 0.22$. (b) Factor of thermal conductivity suppression by rough
edges $r$ versus temperature for ribbon length $N=600$ (line 5) and  $N=400$ (line 6).
}
\label{fig09}
\end{figure}

We consider a ribbon which consists of $K=12$ molecular chains. To model the roughness of the
ribbon edges, we delete with probability $p=1-d$ some atoms from the chains with numbers $k=1,...,K_1$
and $k=K-K_1+1,...,K$. Here $K_1$ is a width of the rough edges, $0\le d\le 1$ is their atomic
density and the parameter $p$ determines in fact the {\it porosity} of the ribbon lattice in the
defect edges. We take in the following $K_1=4$ and $p=0.05$.
We computed the thermal conductivity $\kappa(N,T)$ for $N=400$ ($N_c=200$)
and $N=600$ ($N_c=400$), see Fig.~\ref{fig09}.
As one can see from this figure, at high (dimensionless) temperature the ribbon with the
length $N=600$ has
lower thermal conductivity than the ribbon with length $N=400$. This means that the ribbon
behaves much like thermal insulator, in which the thermal conductivity decreases with
the length of the system, see also \cite{kosevich2}. But the situation is changed for low
temperature, for $T<0.18$,  when the longer ribbon has the higher thermal conductivity,
as in the case of perfect nanoribbons, see the inset in Fig.~\ref{fig09}~(a).

In this connection we can define the factor of thermal conductivity suppression by rough edges
$r(T,N)=\kappa_1(N,T)/\kappa_0(N,T)$, where $\kappa_0(N,T)$ is thermal conductivity of a ribbon
with ideal (atomically smooth) edges and length $N$, $\kappa_1(N,T)$ is thermal conductivity
of the rough-edge ribbon with the same length and width. As one can see from Fig.~\ref{fig08}~(b),
the rough edges suppress the thermal conductivity for all temperatures, $r<1$ for $T>0$,
but the suppression monotonously decreases with the decrease of temperature: for
$T\rightarrow 0$ the thermal conductivities of the ideal- and rough-edge ribbons flatten,
$r\rightarrow 1$. This means that long wave acoustic phonons are not
scattered by surface roughness and
therefore at low enough temperature the rough-edge quantum ribbon becomes an ideal (ballistic)
thermal conductor, in which the thermal conductivity increases with the conductor length.
This in turn means that at low enough temperature we approach the limit of ballistic {\it quantum thermal
transport}, when the value of thermal conductance $G_{th}=J/\Delta T$, the ratio of the thermal
flux $J$ through the quasi-one-dimensional thermal conductor to the temperature difference
$\Delta T= T_+-T_-$, has a quantized value $\pi^2k_B^2T/3h$ (per each of the four massless
acoustic modes of the quasi-one-dimensional waveguide), which does not depend on the material
properties (and perfectness) of the thermal conductor, see, e.g.,  Refs. \cite{maynard,rego,schwab}.
This property of low-temperature thermal transport is in drastic contrast to that in the classical
high-temperature regime, in which the same quasi-one-dimensional rough-edge nanoribbon behaves
as a thermal insulator, in which the thermal conductivity decreases with the insulator length, see
Ref. \cite{kosevich2}.

\section{Modeling of heat transport in carbon nanotube \label{s6}}

The computation of the thermal conductivity can be performed by two methods.
In the first method, the system is thermalized with the use of equilibrium molecular dynamics
and then the interaction with the thermostat is switched off. The current-current correlation
function in the system is computed and the coefficient of thermal conductivity is found with
the help of Green-Kubo formula based on this correlation function. In the second method,
the method of non-equilibrium dynamics, the direct modeling of thermal transport is performed.
For this purpose, two ends of the quasi-one-dimensional system are embedded in two thermostats
with different temperatures. The stationary flux of energy is computed and the coefficient of
thermal conductivity is determined from the known energy flux, temperature difference and
length of the system.  In the approach based on the non-equilibrium dynamics, the correct use of
thermostats is important. The Langevin thermostat can be used in such approach.

The both methods model the heat transport within the classical Newtonian dynamics and
therefore do not allow to take into account the quantum effects. These methods can be justified
only for high enough temperature when the statistics of the system becomes completely classical.
Nevertheless the method of the non-equilibrium dynamics can be easily adapted for the use in
the semi-quantum approach to molecular dynamics modeling of thermal transport.
Below we demonstrate this by example of a carbon nanotube.

For the classical modeling of heat transport, we consider finite nanotube with chirality
index (5,5) (see Fig.~\ref{fig04}), and embed its left end in Langevin thermostat with temperature
$T_+=1.1T$ and the right end in a thermostat with temperature $T_-=0.9T$, where $T$ is a
temperature at which the thermal conductivity will be determined. For this purpose the dynamics
of the atoms at the left (right) end of the nanotube should be described by the Langevin equations
(\ref{f1}) with a white noise with the correlation function given by Eq. (\ref{f5})
with temperature $T=T_+$ ($T_-$). Dynamics
of the atoms in the central part of the nanotube we describe with the Hamilton equations
\begin{equation}
M_n\ddot{\rm\bf r}_n=-\partial H/\partial{\rm\bf r}_n.
\label{f23}
\end{equation}
This allows us to find the distribution of temperature $T(x)$ and energy flux $J(x)$ along the nanotube,
see Fig.~\ref{fig10}. The detailed calculation of these quantities can be found in Ref. \cite{savin1}.
In the classical molecular dynamics description, temperature of a particle
can be determined through its average double kinetic energy per mode (or per degree of freedom) as
\begin{equation}
T=M\langle \dot{x}^2+\dot{y}^2+\dot{z}^2\rangle/3k_B\equiv e_{kin}/k_B,
\label{f23a}
\end{equation}
where $M$ and $(x,y,z)$
are particle mass and cartesian coordinates.
\begin{figure}[tbp]
\includegraphics[angle=0, width=1\linewidth]{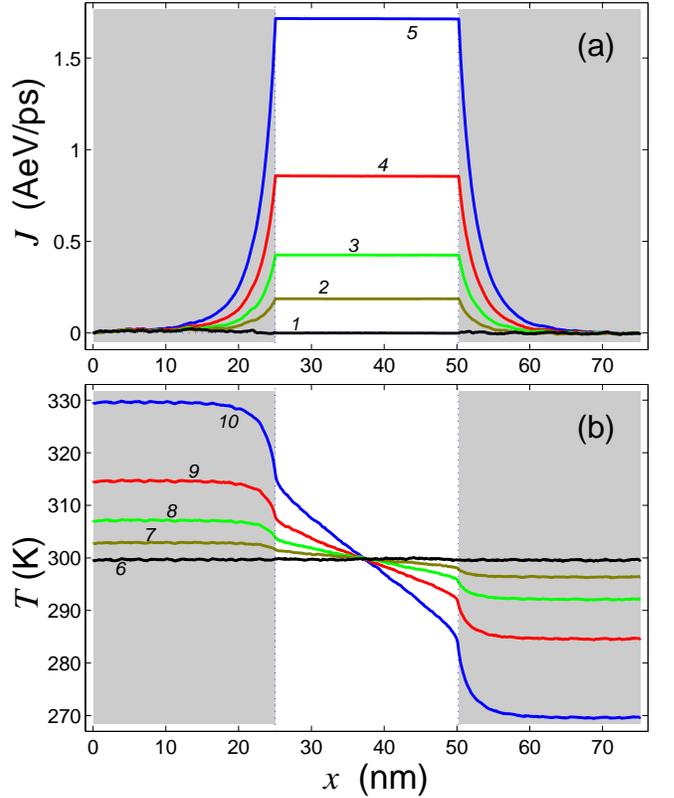}
\caption{(Color online)
Distribution of (a) local energy flux $J$  and (b) local temperature $T$
along the carbon nanotube (5,5) C$_{6020}$, $x$ axis is along the nanotube axis.
Nanotube length is $L=75.04$ nm.
Gray arias indicate the nanotube ends, embedded in classical Langevin thermostats with temperatures
$T_\pm=(1\pm\delta)300$K, for $\delta=0$; 0.0125; 0.025; 0.05; 0.1
(curves 1 and 6; 2 and 7; 3 and 8; 4 and 9; 5 and 10, respectively).
}
\label{fig10}
\end{figure}

To illustrate the modeling of heat transport, we consider a nanotube with a fixed length
$L=75.04$ nm (which corresponds to N=6020 atoms) and embed its ends with lengths $L_e=L/3$ in
classical Langevin thermostats with temperature  $T_\pm=(1\pm \delta)300$ K, where
$\delta=0$, 0.0125, 0.025, 0.05, 0.1. For the relaxation time, we take $t_r=0.4$ ps.
As  one can see in Fig.~\ref{fig10}, for $\delta=0$ when the ends have the same temperature
$T_+=T_-= 300$ K, the temperature is a constant along the whole nanotube ($T(x)\equiv 300$ K)
and there is no heat flux in the circuit ($J(x)\equiv 0$). For $T_+>T_-$ ($\delta>0$), the linear
temperature gradient $T(x)$ and heat flux are formed in the central part of the nanotube:
$J(x)\equiv J>0$ for $L_e\le x\le L-L_e$.
This allows to determine the thermal conductivity of the nanotube of length $L$ as
\begin{equation}
\kappa(L,T)=(L-2L_e)J/S(T_+-T_-),
\label{f24}
\end{equation}
where $S=\pi r^2$ is nanotube cross section [radius $r$ is equal to 0.335 nm for the (5,5)
single-walled nanotube].
One can obtain more accurate estimate for the thermal conductivity from the temperature profile
in the central part of the nanotube:
\begin{equation}
\bar\kappa =(L-2L_e)J/S[T(L_e)-T(L-L_e)],
\label{f24a}
\end{equation}
where $T(x)$ is a distribution of the particle kinetic energy (temperature) along the nanotube.
The definition of thermal conductivity, given by Eq. (\ref{f24a}),
allows one to separate boundary resistances from the thermal resistance of
the nanotube by itself. The boundary (Kapitza) resistance causes finite difference between the
temperatures in the bulk of the heat reservoir and just at the interface with the nanotube,
see Fig. \ref{fig10}(b).
Since one has $T(L_e)<T_+$ and $T(L-L_e)>T_-$ at the interfaces between the ends embedded
in the thermostats and the central part of the nanotube, Eqs. (\ref{f24a})
always give higher
values of thermal conductivity than Eq. (\ref{f24}) does: $\bar\kappa>\kappa$.
Dependencies of the obtained values of $\kappa$ and $\bar\kappa$ on the temperature difference
at the ends of the nanotube,
$\Delta T=2\delta\times 300$ K, are presented in Table \ref{tab2}.
\begin{table*}[tb]
\centering\noindent
\caption{
Heat flux $J$, thermal conductivities $\kappa$ and $\bar\kappa$ of the (5,5) carbon nanotube of
length $L=75.04$~nm versus temperature difference at the ends $\Delta T=T_+-T_-$
[mean temperature is $T=(T_++T_-)/2=300$~K,
lengths of the embedded in thermostats ends are $L_e=L/3$, relaxation time is $t_r=0.4$~ps].
The values were obtained with the use of classical molecular dynamics (CMD)
and of semi-quantum molecular dynamics (SQMD).
The values of $\kappa$ and $\bar\kappa$ have been calculated with the use of Eqs. (\ref{f24}) and (\ref{f24a}).
}
\label{tab2}
\begin{tabular}{c|cc|cc}
\hline
\hline
~    & \multicolumn{2}{|c|}{CMD} & \multicolumn{2}{|c}{SQMD}\\
\hline
~~~$\Delta T$ (K)~~~ & ~~~$J$ (eV\AA/ps)~~~ & ~~~$\kappa$ $(\bar\kappa)$ (W/mK)~~~ &  ~~~$J$ (eV\AA/ps)~~~ & ~~~$\kappa$ $(\bar\kappa)$ (W/mK)~~~ \\
\hline
60.0 &  1.714  & 260.0 (448) & 0.659  & 99.9 (260) \\
30.0 &  0.856  & 259.5 (447) & 0.315  & 95.4 (249) \\
15.0 &  0.426  & 258.6 (443) & 0.154  & 93.4 (244) \\
~7.5 &  0.210  & 257.1 (436) & 0.077  & 93.0 (243) \\
\hline
\hline
\end{tabular}
\end{table*}

As one can see in Table~\ref{tab2}, for $\Delta T\le 60$~K the halving of the temperature
difference results in the halving of the energy flux $J$.
Since in this case the flux is proportional to the
temperature difference, the determination of the thermal conductivity
$\kappa$ ($\bar\kappa$) does not depend on the value of $\Delta T$. It is worth mentioning
that the accuracy of the determination of the energy flux
$J$ decreases with the decrease of $\Delta T$. Therefore the most convenient for the MD modeling
of thermal conductivity are the values of the temperatures at the nanotube ends as $T_+=1.1T$
and $T_-=0.9T$, although a relatively high gradient of temperature is formed in this case
(0.8~K/nm for $T=300$~K).
\begin{figure}[tbp]
\includegraphics[angle=0, width=1\linewidth]{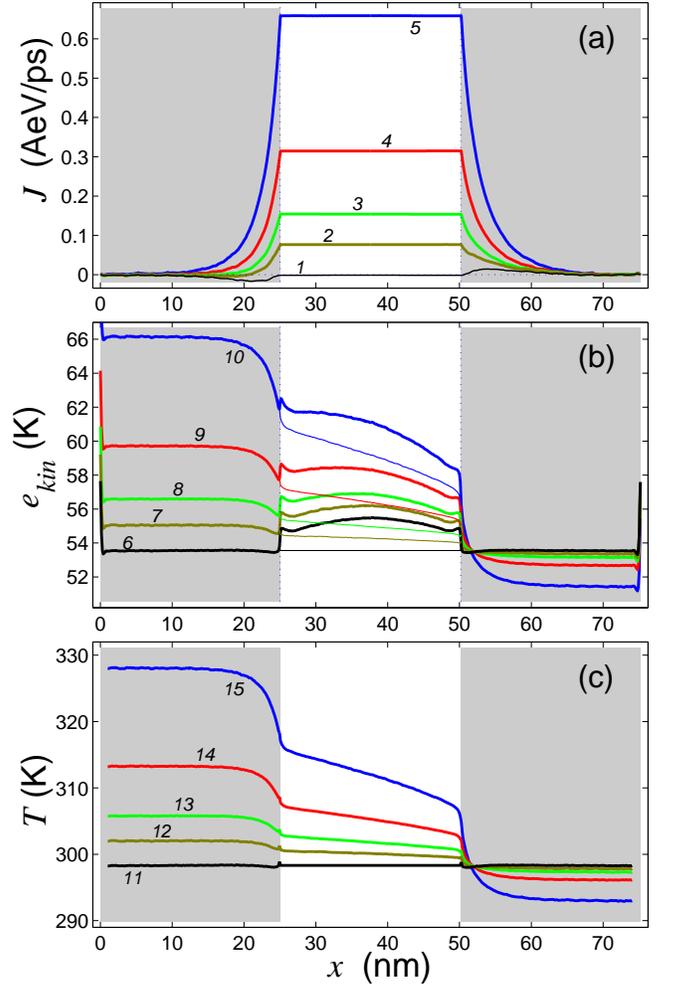}
\caption{(Color online)
Distribution of (a) local energy flux $J(x)$, (b) local kinetic energy per mode $e_{kin}(x)$,
and (c) local temperature $T(x)$
along the carbon nanotube (5,5) C$_{6020}$ axis $x$. Nanotube length is $L=75.04$ nm.
Gray arias indicate the nanotube ends, embedded in  semi-quantum Langevin thermostats with temperatures
$T_\pm=(1\pm\delta)300$K, for $\delta=0$; 0.0125; 0.025; 0.05; 0.1, curves 1 and 6; 2 and 7;
3 and 8; 4 and 9; 5 and 10, respectively. Thin lines show distributions of average particle energies $\bar{e}_{kin}(x)$
in the central part of the nanotube without the artefact surplus thermal energy $\Delta e_{kin}(x)$.
Local temperature $T(x)$ in (c) is determined from Eq. (\ref{sa6}) with the use of  $\bar{e}_{kin}(x)$, given by thin lines in (b).
}
\label{fig11}
\end{figure}

For the semi-quantum MD modeling of heat transport, we must use the Langevin equations (\ref{f1})
with the random forces $\xi_{\alpha n}$ with the power spectral density, given by
$\langle\xi_{\alpha n}\xi_{\beta m}\rangle_\omega
=2M\Gamma k_BT\langle S_{1\alpha n}S_{1\beta m}\rangle_{\bar\omega}$, see Eq. (\ref{f20}),
with temperature $T=T_+$ or $T=T_-$. The color noise is determined as
$\xi_{\alpha n}(t)=\xi_{\alpha n}(\hbar\tau/k_BT_\pm)=k_BT_\pm\sqrt{2M\Gamma/\hbar}S_{1\alpha n}(\tau)$.
The zero-point oscillations do not contribute
to the thermal transport, and therefore they will not be taken into account in the following
and in all the formulas below we will put $S_{0\alpha n}(\tau)\equiv 0$.
The thermal conductivity $\kappa$ can be found with the use of Eq. (\ref{f24}).

In the semi-quantum MD modeling, the mean value of particle kinetic energy $e_{kin}$ does not coincide
with the temperature. This value depends not only on the temperature, but also on the density of
phonon states. For example, the carbon atoms at the end hemispheres of the nanotube have different vibrational
spectrum than the atoms in the central part of the nanotube.
As one can see in Fig.~\ref{fig11}~(b), at the very ends of the nanotube, for $x$ close to 0
and to 75~nm, the mean kinetic energy of carbon atoms is higher than that in the central part of the
nanotube. In the central part of the nanotube, all the atoms have the same vibrational spectrum.
Here $e_{kin}$ depends unambiguously on temperature and therefore the distribution of kinetic energy
along the nanotube characterizes unambiguously the distribution of temperature.

Carbon nanotube is a stiff molecular system with weakly nonlinear dynamics. One can compute its
vibrational spectrum and then use Eq. (\ref{sa6}) to determine the local temperature in the system.

We consider the nanotube of length $L=75.04$ nm, which ends with the length $L_e=L/3$ are embedded into
semi-quantum Langevin thermostats with temperatures $T_\pm=(1\pm \delta)300$~K
(when $\delta=0$, 0.0125, 0.025, 0.05, 0.1).
As one can see in Fig.~\ref{fig11}~(b), for $T_+=T_-=300$~K ($\delta=0$) the mean particle kinetic
energy $e_{kin}$ is not constant along the nanotube. In the central part, which does not interact
directly with the thermostats, particles have higher thermal energy than in the end parts,
which do interact  directly with the thermostats.
For the end parts the mean kinetic energy is $e_{kin}(T_\pm)=53.5$~K, while in the
central part one has $e_{kin}(T_\pm)\le e_{kin}(x)< 55.5$~K. The surplus kinetic energy
$\Delta e_{kin}(x)=e_{kin}(x)-e_{kin}(T_\pm)$, which is $0\le \Delta e_{kin}(x)<2K$, is an artefact
of the semi-quantum molecular dynamics approach when the latter is applied to nonlinear lattice, but it does not produce any thermal flux:
$J(x)=0$ for $L_e<x<L-L_e$, see Fig.~\ref{fig11}~(a).

The appearance of the surplus kinetic energy in the nonlinear system is related with phonon interaction caused by the anharmonicity,
which induces the transfer of energy from the low-frequency modes to the high-frequency ones.
In the nanotube ends, such transfer is suppressed by viscous friction, but in the central part of
the nanotube it can result in the surplus excitation of the high-frequency modes.
In the case of the white-noise heat baths, such inter-mode transfer does not occur because
of the equipartition excitation.
The inter-mode transfer results in surplus excitation of high-frequency modes in the central part
of the nanotube, which ends are embedded in color-noise thermostats. The analysis of the vibrational
spectrum shows that weakly anharmonic interparticle potential in the carbon nanotube results
in a slight accumulation of energy in the high-frequency modes, with $\omega>900$ cm$^{-1}$,
while the low-frequency modes, with $\omega <900$ cm$^{-1}$, follow the Bose-Einstein distribution,
see Fig. \ref{fig12}. As it was explained  above in connection with the quantum definition of the lattice
temperature, see Eqs. (\ref{f035}) and (\ref{f038}), this means that the high-frequency modes possess higher effective
temperature (and therefore are "hot phonons") then the lattice temperature, which is determined in turn
by the Bose-Einstein distribution of the low-frequency modes.

For the system with the harmonic interparticle potential, the Bose-Einstein distribution is valid
for all phonon modes in all the lattice system with the ends, embedded in color-noise thermostats.
Indeed, as we have shown in Section V, there is no surplus excitation of high-frequency modes in the
central part of the nanoribbon with the $harmonic$ interparticle  potential.

For $T_+>T_-$ ($\delta>0$), a constant heat flux is formed in the central part of the nanotube
($J(x)\equiv J>0$ for $L_e<x<L-L_e$), which is proportional to the temperature difference
$\Delta T=T_+-T_-$, see Table \ref{tab2}.
As one can see in Fig.~\ref{fig11}~(b), the distribution of local kinetic energy per mode
$e_{kin}(x)$ has a nonlinear form for $L_e<x<L-L_e$. But after the subtraction of the artefact surplus contribution
$\Delta e_{kin}(x)$ from the average kinetic energy $e_{kin}(x)$, the remaining local energy $\bar{e}_{kin}(x)$ has a linear slope.
If one determines the local temperature $T(x)$ from the distribution of the local
energy  $\bar{e}_{kin}(x)$ with the use of Eq. (\ref{sa6}), one gets the linear distribution of
local temperature along the tube, see  Fig.~\ref{fig11}~(c). The same linear distribution of local
temperature can also be obtained from Eq. (\ref{sa5}), in which the integration over frequencies
is performed in the low-frequency domain $\omega_1=0 \le\omega<\omega_2=900$~cm$^{-1}$. As one can
see in Fig. ~\ref{fig12}, in this frequency domain the vibrational spectrum has a correct
Bose-Einstein distribution, which allows the correct definition of temperature with the use
of Eq. (\ref{sa6}) in all parts, embedded ends and free central part, of the nanosystem.
The linear distribution of local temperature along the tube allows in turn to determine more
accurately the thermal conductivity of the system
with the use of Eq.~(\ref{f24a}).

As one can see from Table \ref{tab2}, for $\Delta T<60$~K the value of thermal conductivity
changes only weakly with the change of temperature difference. Thus for the numerical modeling,
one can use $T_\pm=(1\pm\delta)T$ with $\delta=0.1$. Such temperature difference allows one
to find rather fast the distribution of the heat flux and temperature along the nanotube, and
the obtained value of the thermal conductivity $\kappa$ differs little from the values obtained
for smaller $\delta$.
\begin{figure}[tbp]
\includegraphics[angle=0, width=1\linewidth]{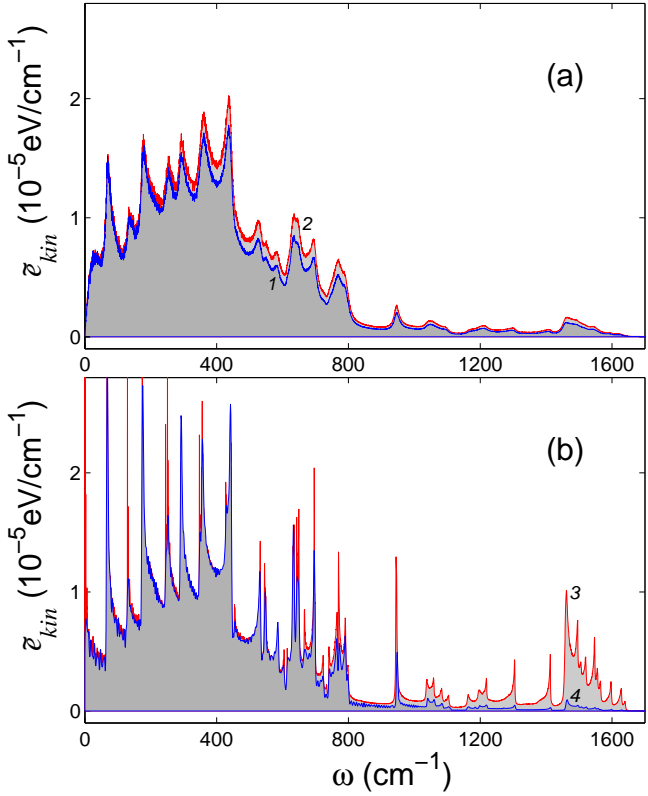}
\caption{(Color online)
Power spectral density of thermal vibrations in carbon nanotube (5,5) with length $L=75.04$~nm
(a) for the atoms at the left end with temperature $T_+=330$~K (curve 2, red), right end with temperature
$T_-=270$~K (curve 1, blue).
(b) Line 3 (red) gives vibrational spectral density in the central part of the nanotube,
which does not interact with the thermostats, line 4 (blue) gives the spectral density given
by Eq. (\ref{sa3}) for the corresponding system of quantum oscillators. The frequency spectrum
of the mean kinetic energy per atom is shown.
}
\label{fig12}
\end{figure}

Now we turn to the analysis of the vibrational frequency spectrum of carbon atoms near the
nanotube ends embedded in thermostats with different temperatures,  $T_+=330$~K and  $T_-=270$~K,
and in the nanotube central part. As one can see in Fig.~\ref{fig12}, the power spectral density
of thermal vibrations at the nanotube ends corresponds to the quantum statistics of phonons.
But there are some distinctions in the nanotube central part, namely the high-frequency modes
with frequencies $\omega>900$~cm$^{-1}$ are  excited more strongly.
This effect is related with the phonon interaction caused by anharmonicity,
which induces the transfer of energy from the low-frequency vibrations to the high-frequency ones.
In the nanotube ends, such transfer is suppressed by viscous friction, but in the central part of
the nanotube it can result in the surplus (artefact) excitation of the high-frequency modes.
As one can see from Fig.~\ref{fig12} (b), this effect is rather weak.
It does not change the heat flux considerably for the used values of the nanotube length.
It is worth mentioning in this connection that the high-frequency modes, which are revealed in
Fig.~\ref{fig12} (b) for  $\omega>900$~cm$^{-1}$, determine, via Eq. (\ref{f035}) or Eq. (\ref{f038}),
the effective temperature of hot phonons, which is higher than the temperature of the equilibrium,
Bose-Einstein, phonons. But the surplus hot-phonon temperature, similar to the surplus kinetic energy
of the particles in the central part of the nanotube, shown in Fig.~\ref{fig11}~(b),
does not contribute to the linear distribution of the temperature and therefore
to the thermal conductivity of the nanotube.

The difference in the power spectral density close to zero frequency in the end, Fig. \ref{fig12} (a),
and central, Fig. \ref{fig12} (b), parts of the nanotube is related with the use of the effective fix-end
and free-end boundary conditions in the simulation of corresponding spectra.

We would like to mention that the obtained results for the temperature dependence
of the specific heat and thermal flux in the considered carbon nanotubes
almost do not depend on the considered relaxation times $t_r=0.1$, 0.2, 0.4, 0.8~ps. The longer
relaxation time increases the computation time. The steady-state heat transport established longer
time and the computation of the mean values requires the integration along longer phase-space
trajectories.  The shorter relaxation time the effective viscosity smears out vibrational
spectra of particles interacting  with the thermostats. It can also increase the reflectivity
of high-frequency phonons at the interfaces between the nanotube central part and the
thermostats. The optimal relaxation time for the carbon nanotube is $t_r=0.4$~ps.

Now we consider the nanotube of the fixed length $L=50.08$ nm (which corresponds to $N=4020$ atoms),
with $L_e=L/4$ and  $t_r=0.4$~ps.
Temperature dependence of the thermal conductivity of such nanotube is shown in Fig.~\ref{fig13}.
As one can see from this figure, within the classical description the thermal conductivity
$\kappa_1$ monotonously increases with the decrease of temperature. This property of
"classical" \ thermal conductivity is related
with the decrease of anharmonicity of lattice dynamics with the decrease of temperature,
which in turn results in the increase of phonon mean free path. But the situation changes
drastically with an account for quantum statistics of phonons. In the semi-quantum description, thermal
conductivity $\kappa_2$ first increases with the decrease of temperature from the high enough
one but then it reaches its maximal value at $T=350$K and monotonously decreases to zero
($\kappa_2\rightarrow 0$ for $T\rightarrow 0$). Such temperature dependence of thermal conductivity,
which is characteristic for solids \cite{peierls,ziman}, is related with the quantum decrease
of the specific heat of the system.
This is confirmed by the property that at low temperature $T\le 250$K the thermal conductivity
of the nanotube in the semi-quantum approach $\kappa_2(T)$ is described with high accuracy by the
relation $\kappa_2(T)\approx c(T)\kappa_1(T)$, where $\kappa_1(T)$ is thermal conductivity of the carbon
nanotube computed within the classical description, $c(T)=C(T)/3Nk_B$ is nanotube dimensionless
specific heat at temperature $T$, see Figs.~\ref{fig05} and \ref{fig13}.
\begin{figure}[tbp]
\includegraphics[angle=0, width=1\linewidth]{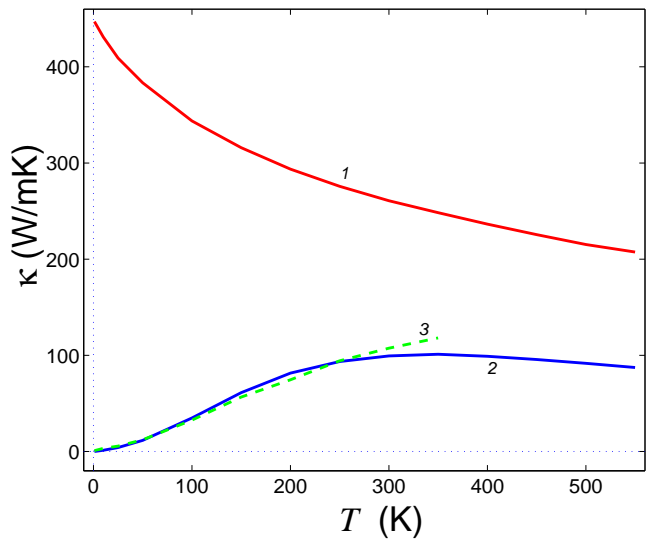}
\caption{(Color online)
Temperature dependence of thermal conductivity $\kappa_i$, $(i=1,2)$,
of the carbon nanotube (5,5) with length $L=50.08$ nm: solid lines 1 and 2 are obtained within
the classical and semi-quantum descriptions, respectively; dashed line 3 gives the temperature
dependence $c(T)\kappa_1(T)$, where $c(T)=C(T)/3Nk_B$ is dimensionless  specific heat
of the carbon nanotube at temperature $T$.
}
\label{fig13}
\end{figure}
\begin{figure}[tbp]
\includegraphics[angle=0, width=1\linewidth]{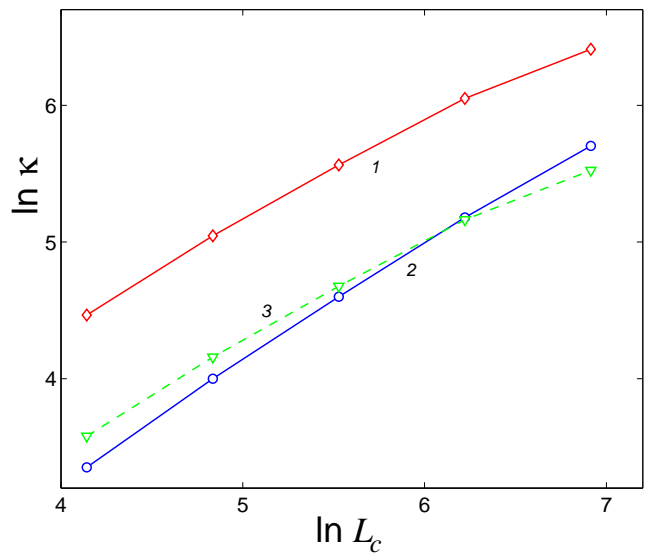}
\caption{(Color online)
Thermal conductivity $\kappa$ versus length of the central part $L_c=L-2L_e$ of the
carbon nanotube (5,5) obtained with the use of classical (solid line 1) and semi-quantum
(solid line 2) descriptions, and thermal conductivity $c(T)\kappa_1(T)$ (dashed line 3,
as line 3 in Fig. \ref{fig11})  at $T=300$K.
The length $L_c$ is measured in \AA, the thermal conductivity $\kappa$ - in W/mK.
}
\label{fig14}
\end{figure}

Length dependence of the nanotube thermal conductivity is shown in Fig.~\ref{fig14}. As one
can see from this figure, both methods give monotonous increase of the thermal conductivity
with the nanotube length. The semi-quantum description gives the lower value of the
conductivity than the classical one for all the considered nanotube lengths.
As one can see from this figure, line 2 lies below line 3 for relatively short nanotubes, which
is consistent with Fig.~\ref{fig13}. But at some nanotube length the line 2 intersects the line 3.
This means that for longer nanotube lengths the mean free path of "semi-quantum" phonons is
larger than that of purely "classical" phonons. This in turn is related with the decrease in
general of phonon mean free path with phonon frequency and the property that the {\it mean frequency}
of semi-quantum phonons is always lower than that of classical phonons. The intersection at some
nanotube length of the line 2 with line 3 in Figs. ~\ref{fig13} and \ref{fig14} means that
short enough nanotubes effectively filter out the high-frequency and short-mean-free-path phonons,
as it occurs in mesoscopic one-dimensional samples, see, e.g., Ref. \cite{maynard}.
Line 2 should intersect the line 1 at even longer lengths, when thermal conductivity of the
"semi-quantum" nanotube will become larger than that of purely "classical" nanotube.
Figure~\ref{fig15} (b) below  shows that similar situation can also be realized in a nanoribbon.
Concerning carbon nanotubes, at $T=300$K such intersection can be
realized only in very long nanotubes because of their very high Debye
temperature (which in fact exceeds the melting temperature of the material).

\section{Thermal conductivity of nanoribbon with periodic interatomic potentials \label{s7}}

In order to demonstrate the characteristic temperature dependence of thermal conductivity of
quantum low-dimensional system with strongly  nonlinear interatomic interactions, we consider a
nanoribbon with periodic interatomic potentials and perfect (atomically smooth) edges,
cf. Eqs.  (\ref{f25})--(\ref{f28}):
\begin{equation}
V(r)=\epsilon_1[1-\cos(r)],~~~U(r)=\epsilon_2[1-\cos(r)]~.
\label{f34}
\end{equation}
In the following we will consider $\epsilon_1=1$, $\epsilon_2=0.5$, when for small relative
displacements of the nearest-neighbor atoms $r$ the potentials (\ref{f34}) will coincide with
the harmonic potentials given by Eq. (\ref{f28}). With the periodic interatomic potentials,
given by Eq. (\ref{f34}),
the ribbon becomes a system of coupled rotators.

A chain of coupled rotators presents a unique translationally-invariant one-dimensional system
with a finite thermal conductivity  \cite{giardina,savin2}. In such system rotobreathers can be excited,
which cause strong {\it momentum-nonconserving} scattering of phonons and result in finite thermal
conductivity of the translational-invariant
system. The density of rotobreathers increases with temperature increase and correspondingly the phonon
mean free path decreases. In the classical description, this causes the monotonous decrease of
thermal conductivity $\kappa$ with the increase of temperature: $\kappa\rightarrow 0$ for
$T\rightarrow\infty$. In the opposite limit $T\rightarrow 0$, the phonon mean free path
monotonously increases and therefore
the thermal conductivity diverges: $\kappa\rightarrow \infty$ for $T\rightarrow 0$.

We consider a ribbon with a width $K=2$ and length $N=800$. We embed the ribbon ends with length
$N_e=300$ each into Langevin thermostats with the color noise with temperature
$T_+=1.1T$ and $T_-=0.9T$ in the left and right edge, respectively. For the relaxation time,
we take $t_r=100$.
Integration of Eqs. (\ref{f30}) for $n=1,...,N_e$ and  $n=N-N_e+1,...,N$,
and of frictionless equations, Eqs. (\ref{f30}) with $\tilde\Gamma=0$ and $\eta_{kn}$, for $n=N_e+1,...,N-N_e$,
allows us to find an average energy flux along the ribbon $J$. With the use of Eq. (\ref{f33}), we
then obtain the thermal conductivity of the finite-length nanoribbon.
\begin{figure}[tbp]
\includegraphics[angle=0, width=.95\linewidth]{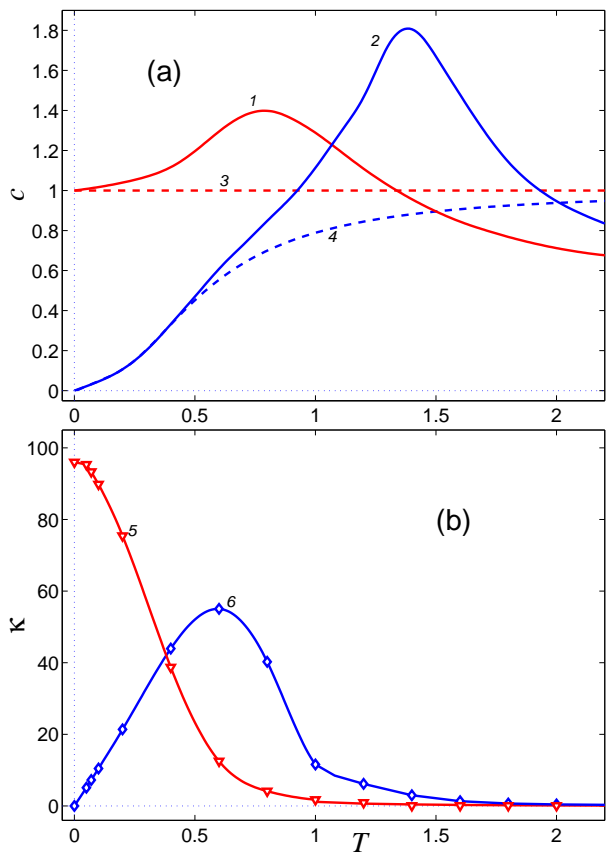}
\caption{(Color online)
(a) Specific heat $c$ of an ideal-edge ribbon with periodic interatomic potential
(\ref{f34}) versus temperature $T$, obtained within the classical (line 1) and semi-quantum
(line 2) descriptions. Dashed lines 3 and 4 show the dependencies for the nanoribbon with the
harmonic interatomic potential (\ref{f28}) within, respectively, the classical and semi-quantum descriptions.
(b) Thermal conductivity $\kappa$ versus temperature $T$ for the ribbon with length $N=800$ and
width $K=2$
(the length of each of the two ends embedded into the thermostats with different temperatures is $N_e=300$):
lines 5 and 6 give the results obtained within the classical and semi-quantum descriptions, respectively.
}
\label{fig15}
\end{figure}

Results of numerical modeling are presented in Fig.~\ref{fig15}. As one can see from the temperature
dependence of the nanoribbon specific heat, Fig.~\ref{fig15}~(a), within the semi-quantum description
the anharmonicity of atomic dynamics starts to show up for $T>0.5$. Just at this temperature specific
heat of the ribbon with the nonlinear interatomic potentials
(\ref{f34}) starts to deviate from the specific heat of the ribbon with harmonic potentials
(\ref{f28}), see Fig.~\ref{fig15}~(a), line 2. It is worth mentioning that in the classical
description the anharmonicity of lattice dynamics starts to show up for lower temperatures
than that in the semi-quantum approach. This is related with the property that the anharmonicity
is more pronounced in the dynamics of short wave phonons and therefore the quantum freezing
out of the high frequency oscillations results in the effective decreasing of the dynamics
anharmonicity (nonlinearity) at low temperature.

In the classical description, the thermal conductivity of the ribbon with nonlinear interatomic
potentials (\ref{f34}) decreases monotonously with the increase in temperature:
$\kappa\rightarrow 0$ for $T\rightarrow\infty$, see Fig.~\ref{fig15}~(b), line 5. Within the
semi-quantum description, the ribbon thermal conductivity reaches its maximal value at $T=0.6$.
For the lower and higher temperatures, the thermal conductivity decreases, $\kappa\rightarrow 0$ for
$T\rightarrow 0$ and $\kappa\rightarrow 0$ for $T\rightarrow\infty$, see Fig.~\ref{fig15}~(b),
line 6. For the low temperature $T<0.5$, system dynamics becomes almost linear when phonons propagate
along the ribbon almost ballistically and the decrease of thermal conductivity is related with
the quantum decrease of the specific heat. In this low temperature limit the thermal conductance
of a short enough ribbon, in which phonon mean free path is longer than the ribbon length, can
reach the lowest (quantum) value which has a linear temperature dependence, see line 6 in
Fig.~\ref{fig15}~(b) and compare the corresponding limit for the rough-edge ribbons,
shown in Fig.~\ref{fig09}, (a) and (b).

For temperature $T>0.4$, the semi-quantum description gives higher value for the thermal
conductivity than that of the classical description. This is also related with the quantum freezing
out of the high-frequency oscillations, which results in the decrease of the mean frequency of
thermal phonons  and correspondingly in the increase of phonon mean free path and phonon thermal
conductivity. Therefore for $T>0.4$ the effect of the increase of phonon mean free path exceeds
the effect of the decrease of low temperature specific heat. At high temperature,
when phonons can be described with the classical statistics, the thermal conductivities given by
the classical and semi-quantum descriptions merge, but the conductivity given by the semi-quantum
description is always higher because of higher phonon mean free path under the equal, classical
and semi-quantum, specific heats.  The temperature dependence of thermal conductivity, given by
line 6 in Fig.~\ref{fig15}~(b), reminds the known temperature dependence of thermal conductivity
in insulators, when the temperature of maximal thermal conductivity separates the low-temperature
boundary-scattering regime from the high-temperature anharmonic Umklapp-scattering regime,
see, e.g., Refs. \cite{peierls,ziman,klitsner,han}. In the case of the nanoribbon with periodic
interatomic potentials and perfect edges, the maximal thermal conductivity, clearly displayed by
curve 6 in Fig.~\ref{fig15}~(b), is reached for the temperature, at which phonon mean free pass, which is determined in this system by the anharmonic scattering,  levels
off with the ribbon {\it length}. For the higher temperature, the phonon mean free pass becomes
shorter then the ribbon length. Figure \ref{fig15}~(b) presents one
of the main results of this work.

\section{Summary \label{s8}}

In summary, we present a detailed description of semi-quantum molecular dynamics approach in which
the dynamics
of the system is described with the use of classical Newtonian equations of motion while the effects
of phonon quantum statistics are introduced through random Langevin-like forces with a specific
power spectral density (the color noise). We describe the determination of temperature in quantum
lattice systems, to which the equipartition limit is not applied.
We show that one can determine the temperature of such system from the measured power spectrum and
temperature- and relaxation-rate-independent density of vibrational (phonon) states.
We have applied the semi-quantum molecular dynamics approach
to the modeling of thermal properties and heat transport
in different low-dimensional nanostructures. We have simulated specific heat
and heat transport in carbon nanotubes, as well as the heat transport in molecular nanoribbons
with perfect (atomically smooth) and rough (porous) edges, and in nanoribbons with strongly anharmonic
periodic interatomic potentials. We have shown that the effects of quantum statistics
of phonons are essential for the carbon nanotube in the whole
temperature range $T<500$K, in which the values of the specific heat and thermal
conductivity of the nanotube are considerably less
than that obtained within the description based on classical statistics of phonons.
This conclusion is also applicable to other carbon-based materials and systems with high Debye
temperature like graphene, graphene nanoribbons, fullerene, diamond, diamond nanowires etc.
We have shown that quantum statistics of phonons and porosity of
edge layers dramatically change low temperature thermal conductivity of molecular nanoribbons in
comparison with that of nanoribbons with perfect edges and classical phonon dynamics and statistics.
The semi-quantum molecular dynamics approach has allowed us to model the transition in the rough-edge nanoribbons from the thermal-insulator-like behavior at high temperature, when the thermal conductivity decreases with the conductor length, see Ref. \cite{kosevich2},  to the ballistic-conductor-like behavior at low temperature, when the thermal conductivity increases with the conductor length.
We have also shown  that the combination of strong nonlinearity of the interatomic
potentials with quantum statistics of phonons changes drastically the low-temperature thermal conductivity
of the system. The thermal conductivity in such samples demonstrates very pronounced non-monotonous
temperature dependence, when the temperature  of maximal thermal conductivity separates the low-temperature
ballistic phonon conductivity from the high-temperature anharmonic-scattering one. At the temperature
of maximal thermal conductivity, phonon mean free pass levels off with the length of the perfect-edge anharmonic
quasi-one-dimensional system. Such non-monotonous
temperature dependence of thermal conductivity is known in bulk insulators and is very different
from monotonously decreasing with temperature conductivity of nanoribbons with the same interatomic
potentials and classical phonon dynamics and statistics, cf. Refs. \cite{giardina,savin2}.

\section*{Acknowledgements}

A.V. Savin thanks the Joint Supercomputer Center of the Russian Academy
of Sciences for the use of computer facilities. Yu. A. Kosevich and A. Cantarero thank the Spanish
Ministry of Economy and Competitivity for the financial support through grant No. CSD2010-0044,
and the University of Valencia for the use of the computers Tirant and Lluis Vives,
from the Red Espa\~{n}ola de Supercomputaci\'{o}n.
Yu. A. Kosevich acknowledges the University of Valencia for hospitality.

\end{document}